\documentclass[structabstract]{aa}
\usepackage{graphicx}
\usepackage{natbib}
\bibpunct{(}{)}{;}{a}{}{,}
\usepackage{float}

\begin{document}

\title{Massive stars on the verge of exploding: the properties of oxygen sequence Wolf-Rayet stars\thanks{Based on observations obtained at the European Southern Observatory under program IDs 091.C-0934 and 093.D-0591.}}
\author{F. Tramper \inst{\ref{api}} 
	\and S.~M. Straal \inst{\ref{api}, \ref{astron}}
	\and D. Sanyal\inst{\ref{bonn}}
	\and H. Sana\inst{\ref{stsci}}
	\and A. de Koter\inst{\ref{api}, \ref{leuven}}
	\and G. Gr\"afener\inst{\ref{armagh}}
	\and N. Langer\inst{\ref{bonn}}
	\and J.~S. Vink\inst{\ref{armagh}}
	\and S.~E. de Mink\inst{\ref{api}}
	\and L. Kaper\inst{\ref{api}}
		}
\institute{Anton Pannekoek Institute for Astronomy, University of Amsterdam, Science Park 904, PO Box 94249, 1090 GE Amsterdam, The Netherlands\label{api} 
	\and ASTRON, The Netherlands Institute for Radio Astronomy, PO Box 2, 7790 AA Dwingeloo, The Netherlands\label{astron}
	\and Argelander Institut f\"ur Astronomie, University of Bonn, Auf dem H\"ugel 71, D-53121, Bonn, Germany\label{bonn}
	\and Space Telescope Science Institute, 3700 San Martin Drive, Baltimore, MD 21218, USA\label{stsci}	
	\and Instituut voor Sterrenkunde, KU Leuven, Celestijnenlaan 200D, 3001 Leuven, Belgium\label{leuven}
	\and Armagh Observatory, College Hill, BT61 9DG Armagh, UK\label{armagh}
		}

\abstract
{Oxygen sequence Wolf-Rayet (WO) stars represent a very rare stage in the evolution of massive stars. Their spectra show strong emission lines of helium-burning products, in particular highly ionized carbon and oxygen. The properties of WO stars can be used to provide unique constraints on the (post-)helium burning evolution of massive stars, as well as their remaining lifetime and the expected properties of their supernovae.}
{We aim to homogeneously analyse the currently known presumed-single WO stars to obtain the key stellar and outflow properties and to constrain their evolutionary state.}
{We use the line-blanketed non-local thermal equilibrium atmosphere code \textsc{cmfgen} to model the X-Shooter spectra of the WO stars and deduce the atmospheric parameters. We calculate dedicated evolutionary models to determine the evolutionary state of the stars.}
{The WO stars have extremely high temperatures that range from 150 kK to 210 kK, and very low surface helium mass fractions that range from 44\% down to 14\%. Their properties can be reproduced by evolutionary models with helium zero-age main sequence masses of $M_{\mathrm{He, ini}} = 15-25 M_{\odot}$ that exhibit a fairly strong (on the order of a few times $10^{-5} M_{\odot} \mathrm{yr}^{-1}$), homogeneous ($f_\mathrm{c} > 0.3$) stellar wind.}
{WO stars represent the final evolutionary stage of stars with estimated initial masses of $M_{\mathrm{ini}} = 40-60 M_{\odot}$. They are post core-helium burning and predicted to explode as type Ic supernovae within a few thousand years. }


\maketitle

\section{Introduction}\label{sec:intro}

\begin{table*}
\centering
\caption{Overview of the known WO stars.}\label{tab:stars}
\begin{tabular}{l l l c c l l}
\hline\hline \\[-8pt]
ID			&  R.A.		& Dec.				& Spectral type\tablefootmark{a} & $Z$	& SIMBAD ID\tablefootmark{b} 				& Other IDs\\
			& (J2000)	& (J2000)			&		&		$(Z_{\odot})$	&			&	\\		
\hline \\[-8pt]
WR102 		& 17:45:47.56	& $-$26:10:26.9	& WO2 & 1	& \object{V* V3893 Sgr}	& [S71d] 4 (Sand 4) \\
WR142		& 20:21:44.35	& $+$37:22:30.6	& WO2&1	& \object{WR 142}		& [S71d] 5 (Sand 5)	\\
WR93b		& 17:32:03.31	& $-$35:04:32.4	& WO3&1	& \object{WR 93b}	\\
BAT99-123	& 05:39:34.31	& $-$68:44:09.1	& WO3&0.5	& \object{Brey 93}	&[S71d] 2 (Sand 2)	\\
LH41-1042	& 05:18:11.01	& $-$69:13:11.3	& WO4&0.5	& \object{[L72] LH 41-1042}	\\
LMC195-1\tablefootmark{c} & 05:18:10.33	& $-$69:13:02.5	& WO2&0.5	& ---	&\\
DR1			& 01:05:01.61	& $+$02:04:20.6	& WO3&0.15	& \object{NAME DR 1 in IC 1613} & [BUG2007] B 17	\\
\hline \\[-8pt]
WR30a		& 10:51:38.93	& $-$60:56:35.2	& WO4 + O5((f))&1	& \object{V* V574 Car}	& [MS70] 4 (MS4) \\
Sk188		& 01:21:04.13	& $-$73:25:03.8	& WO4 + O4 &0.2	& \object{2MASS J01310412-7325038} & [S71d] 1 (Sand 1) \\
\hline
\end{tabular}
\tablefoot{
The upper part of the table shows the (apparently) single WO stars, the bottom part the binaries.\\ 
\tablefoottext{a}{See Section~\ref{subsec:classification} for the spectral classification of the single stars. Binary classifications are from \cite{massey2000} and \cite{moffat1984}.}
\tablefoottext{b}{http://simbad.u-strasbg.fr/simbad/}
\tablefoottext{c}{Not yet listed in Simbad. Designation from \cite{massey2014}.}
}
\end{table*}

The enigmatic oxygen sequence Wolf-Rayet (WO) stars represent a very rare stage in massive star evolution. Their spectra are characterized by strong emission lines of highly ionized carbon and oxygen, and in particular a strong \ion{O}{vi} $\lambda$3811-34 \AA \ emission line. Their emission-line spectra point to dense, outflowing atmospheres. Despite their rarity, the WO stars can provide key information in our understanding of massive star evolution. 

Although the progenitors of hydrogen-free type Ib and Ic supernovae (SNe) have not yet been identified, it is very likely that they are evolved Wolf-Rayet stars \citep[e.g., ][]{yoon2012}. In particular, WO stars are potential progenitors of the helium-deficient type Ic SNe, as they may have a very low helium abundance. If they are rapidly rotating when they explode, they could produce an associated long-duration gamma-ray burst \citep[e.g.,][]{woosley2006}. 

The tell-tale signature of WO stars is their \ion{O}{vi} $\lambda$3811-34 \AA \ emission. This emission was first found in the spectra of the central stars of planetary nebulae. However,  \cite{sanduleak1971} pointed out that five of the stars showing \ion{O}{vi} $\lambda$3811-34 \AA \ have broad, Wolf-Rayet like emission lines and do not appear to have an associated nebula. It was therefore suggested that they are part of the carbon sequence Wolf-Rayet (WC) class. \cite{barlow1982} argued that these WC \ion{O}{vi} stars should be seen as a separate class of Wolf-Rayet stars, and introduced the first WO classification scheme.  

Since their discovery, the WO class has been commonly interpreted as a very short stage of evolution of massive stars covering an initial mass range of approximately $45-60 M_{\odot}$ after the carbon sequence Wolf-Rayet (WC) phase \citep[e.g.,][]{sander2012, langer2012}. In such a scenario, the emission lines of highly ionized oxygen reflect the high oxygen abundance that is expected near the end of core-helium burning. The very high stellar temperature that is needed to produce \ion{O}{vi} emission is expected if WOs are indeed the descendants of WC stars, as the envelope is being stripped by the stellar wind and consecutively hotter layers are revealed. An alternative scenario is that WO stars originate from higher mass progenitors. In this case the stars are hotter and the \ion{O}{vi} emission could purely be an excitation effect, and a high oxygen abundance is not necessarily implied \citep[e.g., ][]{hillier1999}.

If WO stars represent a later stage of evolution compared to WC stars, the surface abundance of carbon and oxygen is expected to be high. If these can be measured, and a good estimate of the stellar mass can be obtained, the WO stars can provide unique observational constraints on the elusive $^{12}\mathrm{C}(\alpha, \gamma)^{16}\mathrm{O}$ thermonuclear reaction rate. This rate is currently only weakly constrained, with an uncertainty on the order of $30\%$ \citep[e.g., ][]{tur2006}.

Currently, nine members of the WO class are known, two of which are in a binary system. Table~\ref{tab:stars} lists the coordinates, names, and metallicities ($Z$) of all these stars. Two WO stars have recently been discovered in the LMC. The first, the WO4 star LH41-1042, was discovered by \cite{neugent2012}. The second one was reported by \cite{massey2014}, and is, remarkably, located only 9\arcsec \ away from LH41-1042. 

As a first step in our effort to investigate the nature of the WO stars,  \citet[][henceforth Paper I]{tramper2013} performed a detailed spectroscopic analysis of DR1. Located in the Local Group dwarf galaxy IC~1613, this star is the lowest metallicity WO star known,  with a metallicity $Z \sim 0.1-0.2 \, Z_{\odot}$ (Paper I). The stellar parameters that were derived confirm the expected very high stellar temperature, close to the helium terminal-age main sequence (He-TAMS). However, the derived surface abundances of helium, oxygen, and carbon are comparable to those found in early WC stars. Thus, if DR1 is representative for the WO stars, these stars may be descendants of higher mass progenitors than WC stars. 

In this paper, we perform a homogeneous spectroscopic analysis of all the remaining (apparently) single WO stars that are known, with the exception of the recently discovered star in the LMC \citep[LMC195-1,][]{massey2014}. The stellar parameters, together with the results from paper I, are used to determine the nature of WO stars. We use dedicated helium-burning models to constrain their evolutionary stage as well as to predict their remaining lifetime.

In the next section, the observations and data reduction are described. Section~\ref{sec:analysis} outlines the modeling of the observed spectra, and the resulting properties are discussed in Section~\ref{sec:properties}. These are used to determine the remaining lifetime in Section~\ref{sec:lifetime}. Finally, we conclude on the nature of the WO stars in Section~\ref{sec:nature}.

\section{Observations and data reduction}\label{sec:obs}

All observations presented in this work have been obtained at the European Southern Observatory using the X-Shooter instrument \citep{vernet2011} on the {\it Very Large Telescope}. WR142 and WR30a were observed under program ID 093.D-0591, and all the other stars as part of the NOVA program for guaranteed time observations under program ID 091.C-0934. X-Shooter covers a wavelength range from 3\,000 \AA \ to 25\,000 \AA \ by directing the light in three separate arms: the UVB (3\,000-5\,500 \AA), VIS (5\,500-10\,000 \AA), and NIR (10\,000-25\,000 \AA) arms.
\par To prevent detector saturation by the strong emission lines, the observations were split up into several shorter exposures. An overview of the exposure times and slit widths used is given in Table~\ref{tab:observations}. All observations were done in nodding mode with a nod throw of 5\arcsec. The slit was oriented along the parallactic angle for all observations.
\par The data of all stars have been reduced using the X-Shooter pipeline v2.2.0, which produces flux-calibrated 1D spectra. The flux calibration is performed using observations of the spectro-photometric standard stars listed in Table~\ref{tab:observations} taken during the same night. Two of the stars (WR30a and LH41-1042) had another bright object in the slit, and the standard reduction was not sufficient. For these stars, we first obtained the 2D spectrum for each nodding position separately, without sky subtraction. We then subtracted the sky background using a clean part of the slit. 

\begin{table*}
\centering
\caption{Overview of observations.}\label{tab:observations}
\begin{tabular}{l c c c c l}
\hline\hline \\[-8pt]
ID			 & MJD		& $T_{\mathrm{exp}}$\tablefootmark{a} & Slit width\tablefootmark{a} 	& Spec. Standard & Airmass\\
			& {\it At start exp.}					& $(s)$ 				& $(\arcsec)$		&  \\		
\hline \\[-8pt]
WR102 		& 56490.069	& $4 \times 120$				& 0.8, 0.9, 0.9	& GD153	&	1.0\\
WR142 		& 56783.371	& $8 \times 200/260/260$			& 0.8, 0.9, 0.9 	& LTT7987&	2.3\\
WR93b 		& 56489.138	& $4 \times 600$				& 0.8, 0.9, 0.9 	& EG274&	1.0\\
BAT99-123 		& 56409.978	& $10 \times 180$				& 0.8, 0.9, 0.9 	& LTT3218&	1.7\\
LH41-1042		& 56522.351		& $10 \times 180$				& 0.8, 0.9, 0.9	& FEIGE-110&	1.8\\
\hline \\[-8pt]
WR30a 		& 56771.212	& $8 \times 200/260/260$		& 0.8, 0.9, 0.9	& LTT7987 &	1.6\\
SK188 		& 56522.388	& $8 \times 40$					& 0.5, 0.7, 0.4	& EG274&	1.5\\
\hline
\end{tabular}
\tablefoot{
\tablefoottext{a}{If a single value is given it is the same for all three X-Shooter arms; if three values are given it is for the UVB, VIS and NIR arms, respectively.}}
\end{table*}

\par
\begin{figure}
   	\resizebox{\hsize}{!}{\includegraphics{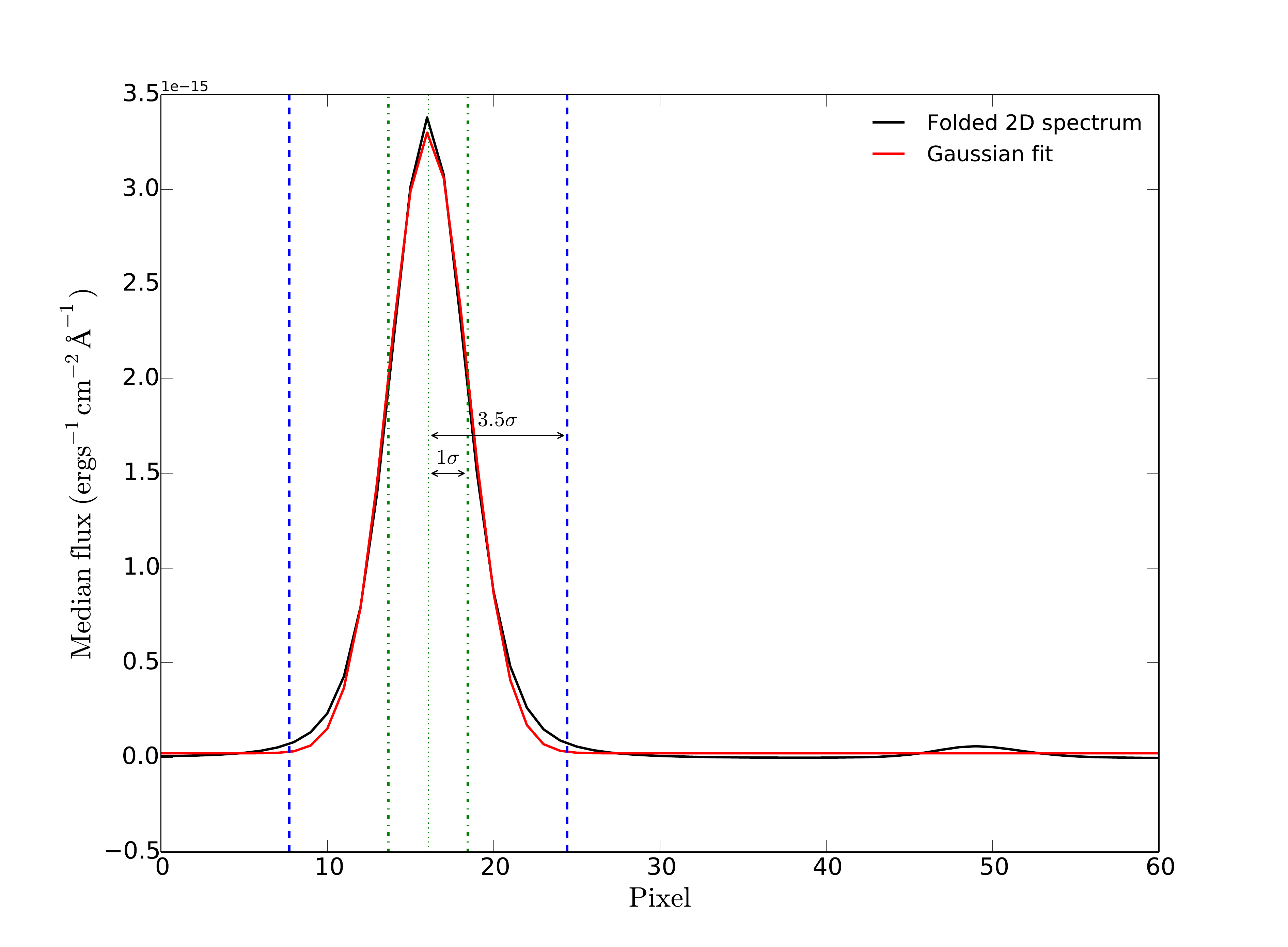}}
  	\caption{Example of the extraction of the 1D spectrum of WR30a for one of the nodding positions. The cross-dispersed profile is shown in black, the Gaussian fit in red. The blue dashed lines indicate the region that is extracted (corresponding to $\pm3.5 \sigma$, with $\sigma$ the standard deviation of the fitted Gaussian profile).}
  	\label{fig:extract}
\end{figure}

 Because the atmospheric dispersion corrector was unavailable at the time of observations, the traces of the two stars mentioned above were not at a constant position on the slit as a function of wavelength in the UVB spectra. We therefore extracted the 1D spectra by integrating the flux at each wavelength over a sufficiently large part of the slit. While this introduces additional noise, the quality of the data is high enough for this to be negligible. In the VIS arm the traces of both stars were at a constant position on the slit. Here, we folded the spectra in the wavelength direction, and fitted a Gaussian to the stellar signal. The 1D spectra were then subtracted using the parts of the slit corresponding to $\pm3.5 \sigma$ covered by the average cross-dispersed point-spread function. Figure~\ref{fig:extract} shows an example, in which the second object in the slit is located around pixel 50. While the median flux of this object over the full wavelength range is very low, it has a noticeable effect on the extracted spectrum if not taken into account, in particular at the shorter wavelengths. For both stars, the flux of the contaminating object was negligible in the NIR, and the 1D spectra were obtained using the nodding reduction, as this facilitates a better correction for the copious telluric lines. The resulting fluxes connect well with the VIS spectra which indicates that the contribution of the second object can indeed be ignored.
\par
\begin{figure}
   	\resizebox{\hsize}{!}{\includegraphics{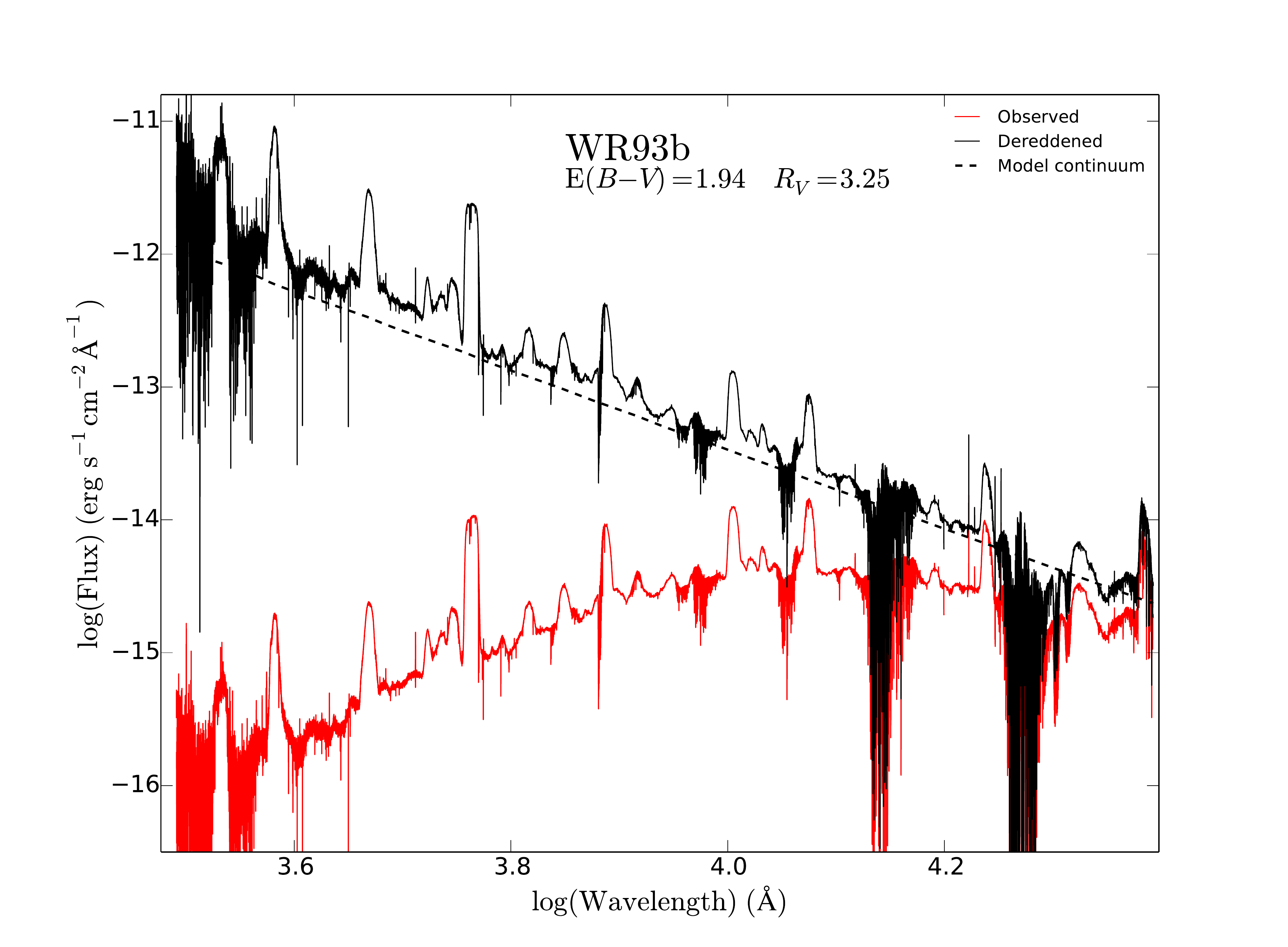}}
  	\caption{Example of the extinction correction procedure for the highly reddened WR93b. The flux-calibrated spectrum before and after the extinction correction is displayed in red and black, respectively. The scaled continuum of a model spectrum is indicated by the dashed black line.}
  	\label{fig:deredden}
\end{figure}

 The flux-calibrated spectra of the stars were extinction corrected using the CCM extinction laws \citep{cardelli1989, odonnell1994}. A {\sc cmfgen} model (see Section~\ref{sec:modeling}) of the corresponding metallicity was used as a template for the slope of the spectrum (see Figure~\ref{fig:deredden} for an example). The value of the total-to-selective extinction $R_V$ was only adjusted if a proper dereddening could not be achieved using the average value of $R_V=3.1$. The derived values for $E(B-V)$ and $R_\mathrm{V}$ are listed in Table~\ref{tab:reddening}. The estimated uncertainty is 10\% for $R_V$ and 0.1 dex for $E(B-V)$.
\par
\begin{figure*}
   	\includegraphics[width=\textwidth]{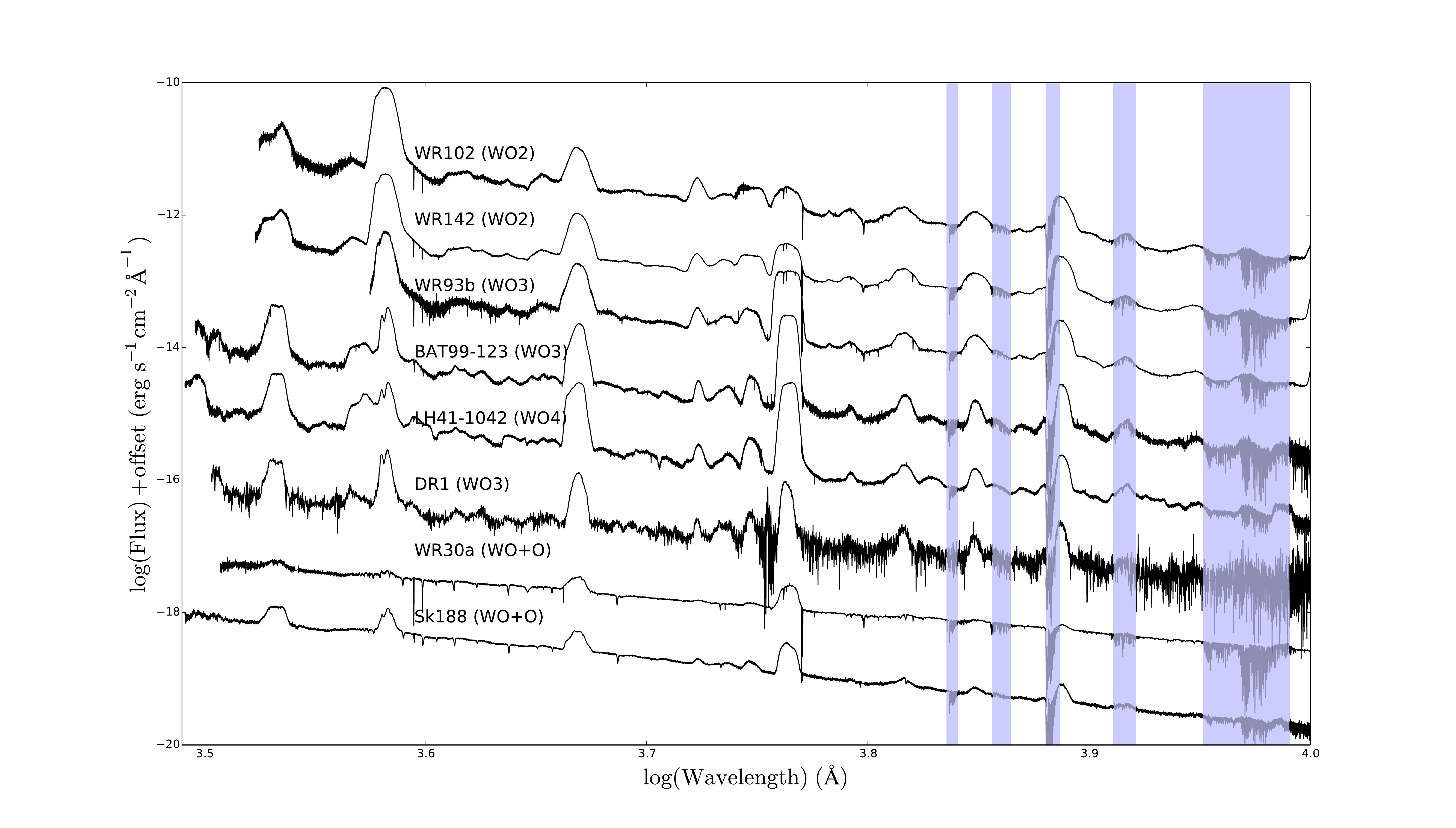}
  	\caption{Dereddened, flux-calibrated X-Shooter spectra in the UVB and VIS range ($3\,000-10\,000$ \AA). The flux has been multiplied by an arbitrary factor for plotting purposes. Nebular emission and residuals from the sky subtraction have been clipped, and wavelength ranges affected by strong atmospheric features are displayed within the shaded areas. The spectrum of DR1 has been rebinned to 1 \AA. }
  	\label{fig:atlas_UVBVIS}

  	\includegraphics[width=\textwidth]{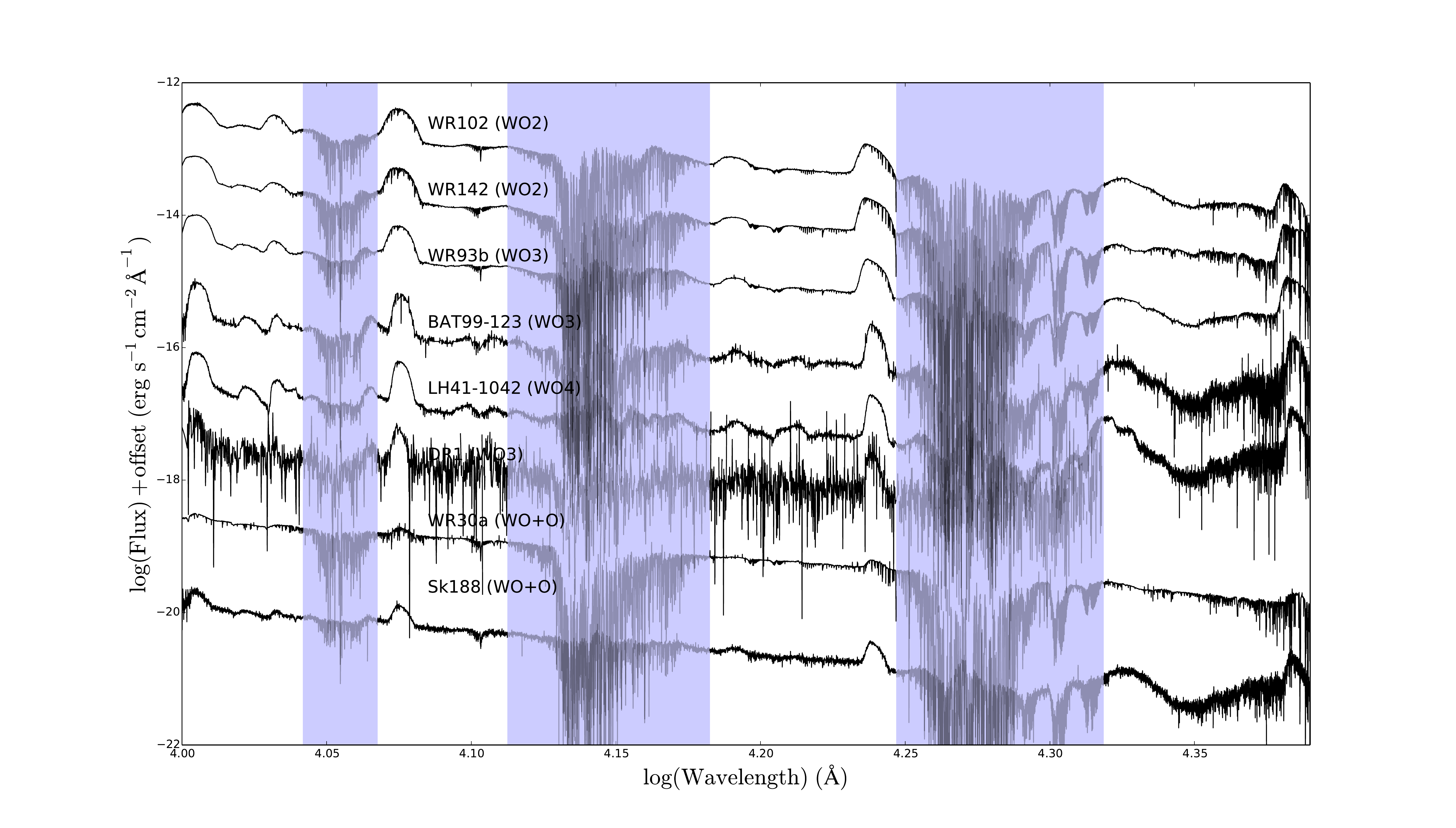}
  	\caption{As Figure~\ref{fig:atlas_UVBVIS}, but for the NIR range ($10\,000-25\,000$ \AA). The spectra of BAT99-123, LH41-1042 and Sk188 have been binned to 1 \AA. The spectrum of DR1 has been binned to 2 \AA, and only extends to 20\,000 \AA \ due to the use of the X-Shooter $K$-band blocking filter.}
  	\label{fig:atlas_NIR}
\end{figure*}

 The spectrum of one of the single WO stars, LH41-1042, shows a steeper slope, but does not display spectral features from a companion. Although a faint star is detected in a UV image at a small projected distance from LH41-1042 (see Figure~\ref{fig:HST}), the spectrum could not be corrected by assuming a Rayleigh-Jeans contribution from this object. Instead, we have artificially corrected the flux to match the model WO slope for LMC metallicity. The drawback of this approach is that the reddening cannot be determined for this object, implying a larger uncertainty in the derived luminosity. The flux correction for LH41-1042 is described in detail in Appendix~\ref{sec:correction_appendix}. The extinction-corrected, flux-calibrated spectra of all WO stars are presented in Figures~\ref{fig:atlas_UVBVIS} and \ref{fig:atlas_NIR}.

\begin{table}
\centering
\caption{Derived reddening and total-to-selective extinction.}\label{tab:reddening}
\begin{tabular}{l c c}
\hline\hline \\[-8pt]
ID			&	$E(B-V)$	& 	$R_V$	\\
\hline \\[-8pt]
WR102		&	1.26		&	3.10		\\
WR142		&	1.72		& 	2.85		\\
WR93b		&	1.94		&	3.25		\\
BAT99-123	&	0.19		&	3.10		\\
LH41-1042	&	---		&	---		\\
\hline
\end{tabular}
\end{table}

\subsection{Spectral classification}\label{subsec:classification}

\begin{figure}
   	\resizebox{\hsize}{!}{\includegraphics{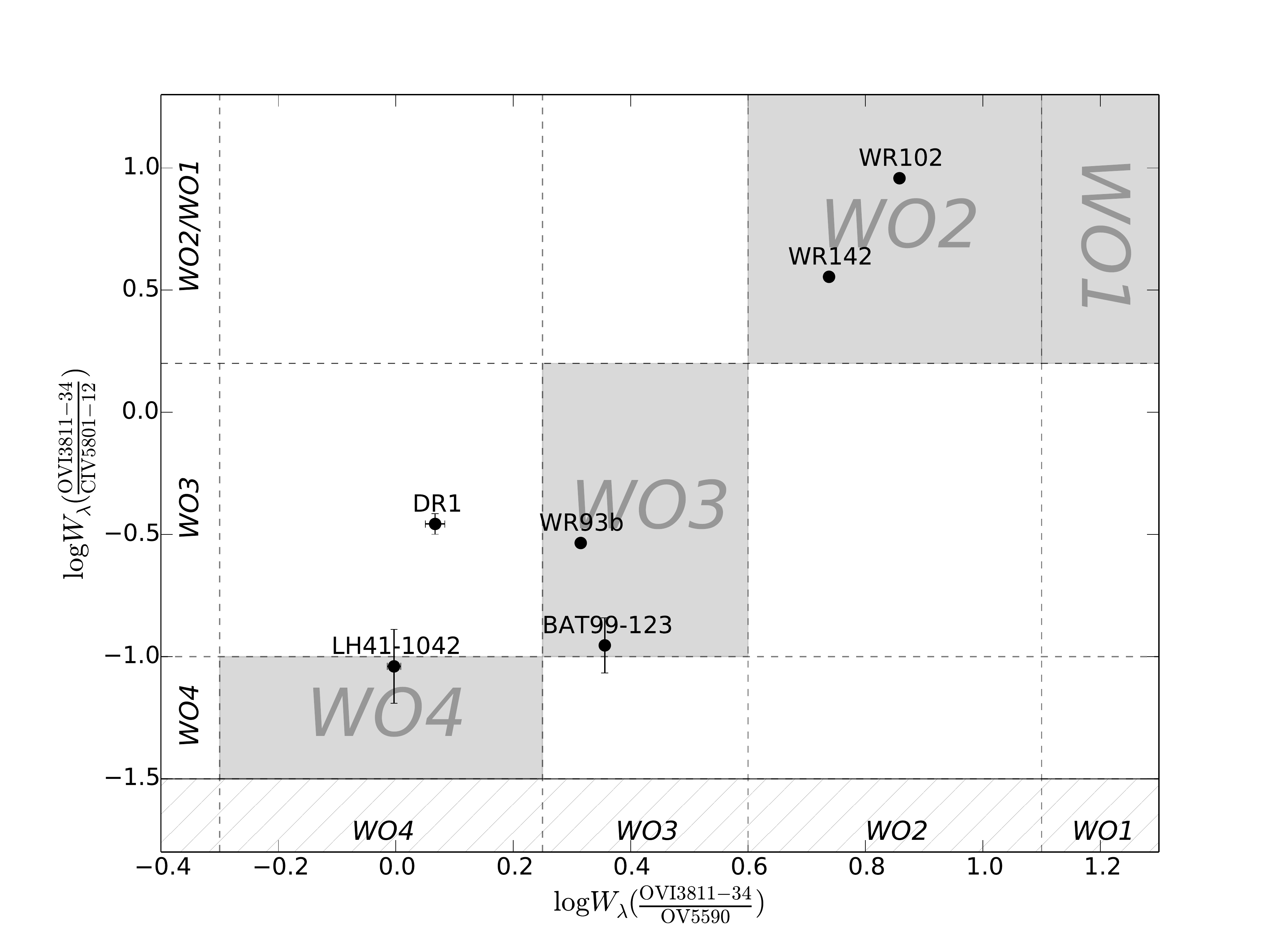}}
  	\caption{Comparison of the primary and secondary spectral classification criteria (x and y axis, respectively). The shaded areas indicate the regions where the two criteria agree on the spectral type. If these are not in agreement, the classification depends on the FWHM of \ion{C}{iv} $\lambda$5801-12 \AA. This is only the case for DR1, which we assign the spectral type WO3, in agreement with previous classifications. The dashed shaded region indicates where the WC sequence begins.}
  	\label{fig:spt}
\end{figure}

Two quantitative classification schemes exist for the WO subclasses. The first divides the WO class in five subclasses, ranging from WO1 to WO5 \citep{kingsburgh1995a}. This classification scheme is based on the equivalent width (EW) ratio of \ion{O}{vi} $\lambda$3811-34 \AA \, to \ion{C}{iv} $\lambda$5801-12 \AA \, and \ion{O}{vi} $\lambda$3811-34 \AA \,  to \ion{O}{v} $\lambda$5590 \AA. 
\par \cite{crowther1998} introduced a classification scheme for WC and WO stars in which the WO class is divided into four subtypes, from WO1 to WO4, and connects to the WC class at WC4. This classification is based on the same EW ratios as the scheme from \cite{kingsburgh1995a}, but also includes the full width at half maximum (FWHM) of \ion{C}{iv} $\lambda$5801-12 and the EW ratio of \ion{O}{vii} $\lambda$5670 to \ion{O}{v} $\lambda$5590. 
\par We adopt the classification scheme of \cite{crowther1998} in this paper. The spectral types that we derive are given in Table~\ref{tab:stars} (see also Appendix~\ref{sec:SpT_appendix}).

\section{Spectroscopic analysis}\label{sec:analysis}

First, we briefly discuss the morphological properties of the flux-calibrated spectra. Following this, we perform a detailed quantitative spectroscopic analysis.

\subsection{Morphological properties}\label{sec:qualitative}

The WO spectra in Figures~\ref{fig:atlas_UVBVIS} and \ref{fig:atlas_NIR} show clear trends with spectral type and metallicity. These trends reflect changes in the physical properties of the star and in the region in the wind where the lines are formed, and provide information for the subsequent modeling (Section~\ref{sec:modeling}). 

Most notable is the increase in line width at higher metallicity. This is visible to a varying degree in all the spectral lines, and can be seen particularly well in the \ion{C}{iv} and \ion{He}{ii} lines in the NIR. The broader lines at higher metallicity reflect an increase in the terminal velocity of the outflows from the stars. 

The morphology of the characterizing \ion{O}{vi} $\lambda$3811-34 \AA \ emission also changes with metallicity. At low metallicities (IC~1613 and LMC, but also for the binary in the SMC) this doublet is clearly double-peaked, while it is fully blended in the spectra of the galactic stars. The \ion{O}{v} $\lambda$5590 \AA \ line profiles change from a roughly parabolic shape in the stars with a sub-galactic metallicity to a broad flat-topped shape in the MW stars. This reflects a change in the optical depth of the line-forming region: the line is formed in the optically thick region of the outflow for the lower metallicities, while for the galactic stars it is formed in optically thin regions.

\subsection{Modeling}\label{sec:modeling}

\begin{table*}
\centering
\caption{Properties of the single WO stars.}\label{tab:properties}
\begin{tabular}{l c c c c c c c c c c c c}
\hline\hline \\[-8pt]
ID			&	$\log{L}$				&	$T_*$	& $R_*$	& $\frac{\mathrm{N}_{\mathrm{C}}}{\mathrm{N}_{\mathrm{He}}}$	&	$\frac{\mathrm{N}_{\mathrm{O}}}{\mathrm{N}_{\mathrm{He}}}$		&	$v_{\infty}$	&	$\log{\dot{M}}$ 		& $\log{R_t}$	&	$\eta$	& $\log{Q_0}$	&	$\log{Q_1}$	& $\log{Q_2}$	\\
			&{\tiny$(L_{\odot})$}	&{\tiny(kK)}	&		{\tiny ($R_{\odot}$)}			&			&								& {\tiny (km s$^{-1}$)}& {\tiny $(M_{\odot}$ yr$^{-1}$)} 	&	{\tiny ($R_{\odot}$)}	&	&	{\tiny (s$^{-1}$)}&	{\tiny (s$^{-1}$)}&	{\tiny (s$^{-1}$)}\\
{\it Typical uncertainty}	& 	&	{\it 20}	&	&	{\it 10\%}		& {\it 10\%}	&	{\it 200}	&	{\it 0.1}	&	\\
\hline \\[-8pt]
WR102		&		5.45$^{+0.15}_{-0.23}$			&	210		& 0.39	&	1.50					&	0.45								&	5000			&		-4.92			&	0.67		& 10.3	&	49.0		&	48.9		&	48.4\\
WR142		&		5.39$^{+0.15}_{-0.23}$			&	200		& 0.40	&	1.00					&	0.16								&	4900			&		-4.94			&	0.79		& 11.3	&	49.0		&	48.8		&	48.3\\
WR93b	&		5.30$^{+0.15}_{-0.23}$			&	160		& 0.58	&	0.60					&	0.15								&	5000			&		-5.00			&	0.97		& 12.3	&	48.9		&	48.8		&	47.3\\
BAT99-123	&		5.20$^{+0.06}_{-0.07}$			&	170		& 0.47	&	0.63					&	0.13								&	3300			&		-5.14			&	0.82		& 7.3		&	48.9		&	48.7		&	47.6	\\
LH41-1042	&		5.26$^{+0.12}_{-0.14}$			&	150		& 0.62	&	0.90					&	0.20								&	3500			&		-5.05			&	0.83		& 8.6		&	48.9		&	48.7		&	45.5	\\
DR1	&		5.68$^{+0.06}_{-0.07}$			&	150		& 1.06	& 	0.35					&	0.06								&		2750		&		-4.76			&	0.46		& 5.0		&	49.5		&	49.3		&	48.0			\\
\hline
\end{tabular}
\end{table*}

To perform a homogeneous quantitative spectroscopic analysis of the WO stars we employ the {\sc cmfgen} code of \cite{hillier1998}. This code iteratively solves the transfer equation in the co-moving frame, and accounts for effects such as clumping and line blanketing. Our fitting strategy has been described in detail in Paper I. Here we only report on relevant assumptions and changes in the applied diagnostics. Our modeling approach aims to reproduce all the observed trends in the sample, as well as to provide a good fit to each of the individual spectra.

\begin{figure}
   	\resizebox{\hsize}{!}{\includegraphics{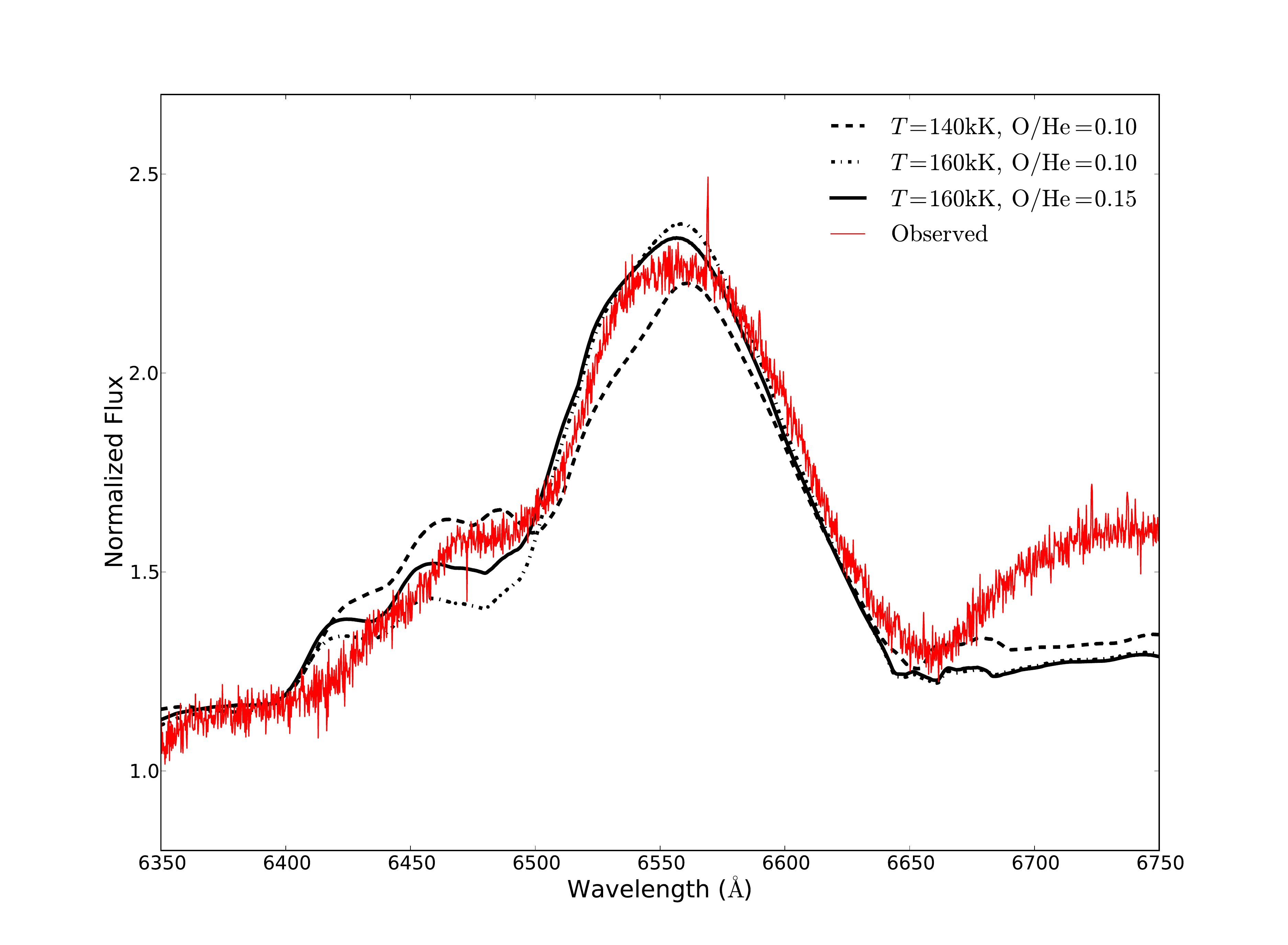}}
  	\caption{Behavior of the \ion{He}{ii}, \ion{O}{v}, and \ion{C}{iv} blend at 6560 \AA \ for different values of the temperature and the oxygen abundance. The observed spectrum is from WR142.}
	\label{fig:temperature}
\end{figure}

In Paper I, the weak optically thin \ion{He}{ii} line at 4859 \AA \ was used as a diagnostic for the stellar temperature. This line is not recognizably present in the spectra of the LMC and MW stars that are analyzed here. Instead, we use the shape of the \ion{He}{ii} $\lambda$6560 \AA \ as a temperature probe. The blue wing of this line has contributions from \ion{O}{v} and \ion{C}{iv}, and the shape of the line profile can only be fitted by models with the correct combination of temperature and carbon and oxygen abundances. Together with the other abundance diagnostics, this allows us to constrain the temperature with an accuracy of about 20 kK (see Figure~\ref{fig:temperature}). 

To determine the luminosities of the stars, we adopt a distance of 50.12 kpc to the LMC \citep{gibson2000}, and distances of 4.6, 1.75, and 3.4 kpc for WO-MW-1, 2, and 3, respectively \citep{drew2004}. We derive spectrophotometric V-band magnitudes for the observed spectrum and compare these to the Simbad values. This gives an estimate of the flux loss and error in the flux calibration of the observations. We model the luminosity by matching the distance-corrected model flux to the magnitude-corrected, dereddened observed flux in the V-band wavelength region. The model flux is not only determined by the luminosity, but is also affected by the mass-loss rate and stellar temperature. Therefore, like the determination of all other parameters, the luminosity determination is an iterative process. 

Table~\ref{tab:properties} presents an overview of the best-fit parameters for all the single WO stars in our sample. A formal error determination is not possible, as we cannot perform a full exploration of parameter space. Table~\ref{tab:properties} gives an estimation of the uncertainty in the derived values, which reflect the sensitivity of the model spectra to changes in that parameter. To estimate the uncertainty in the luminosity, we assume an error of 20\% and 5\% in the distance for the Galactic and extra-Galactic stars, respectively. Based on the comparison with the V-band magnitudes, we estimate an uncertainty of 10\% in the flux calibration. The resulting error bars are given in Table~\ref{tab:properties}. The error on the luminosity of LH41-1042 has been multiplied by two to reflect the absence of the extinction correction (see above).

 Figures showing the best-fit model for each star are given in Appendix~\ref{bestfit_appendix}. Table~\ref{tab:properties} also gives values for the transformed radius $R_t$, the wind efficiency $\eta$, and the ionizing fluxes $Q_{0,1,2}$. The ionizing fluxes indicate the number of photons per second that are available to ionize hydrogen ($Q_0$), and singly ($Q_1$) and doubly ($Q_2$) ionize helium.

The transformed radius, introduced by \cite{schmutz1989}, is defined as:

\begin{equation}
R_t = R_* \left[ \frac{v_{\infty}}{2500 \, \mathrm{km \, s}^{-1}} \middle/ \frac{\dot{M}}{\sqrt{f_{\mathrm{c}}}10^{-4}M_{\odot}\mathrm{yr}^-1} \right]^{2/3},
\end{equation}

\noindent where the volume filling factor $f_{\mathrm{c}}$ was first incorporated by \cite{hamann1998}. The temperature $T_*$ is the temperature at radius $R_*$ at the base of the wind, where the underlying regions can be assumed to be in hydrostatic equilibrium. Combinations of parameters that keep $R_t$ constant produce very similar spectra, making it a very useful quantity. For instance, it implies that when $v_{\infty}$ and $R_*$ are well constrained, the line flux is determined by the value of $\dot{M}/\sqrt{f_c}$. In Section~\ref{sec:lifetime} we use this property to constrain $f_{\mathrm{c}}$.

The wind efficiency parameter gives the ratio of the wind momentum once the flow has reached the terminal velocity ($\dot{M} v_{\infty}$) and the photon momentum ($L/c$):

\begin{equation}
\eta = \frac{\dot{M} v_{\infty} c}{L}.
\end{equation}

\noindent Thus, $\eta$ indicates the average amount of scatterings that photons undergo in order to drive the wind. The values of $\eta$ for the WO stars are in the range of $5\sim12$, very similar to values found for WC stars \citep{sander2012}. Multiple photon scatterings are expected for optically thick winds \citep{de-koter1997, vink2012, grafener2013}. 

Overall, the observed spectra are well reproduced by our models, and allow us to constrain the temperature, surface abundances, and wind properties of the WO stars. Nevertheless, some of the observed spectral features are not fully reproduced. The \ion{O}{vi} $\lambda$3811-34 \AA \ cannot be reproduced while simultaneously fitting the overall spectrum, and the flux in this line is underpredicted by a factor of $\sim3$ in our models. The cause of this is likely to be found in the susceptibility of the population of the upper level of the transition to X-ray excitation (see Paper I).  Still, the strength of this line in our models does follow the observed trend with spectral type.

The observed spectra show some emission lines that are not in the model spectra. These lines belong to transitions of the higher ionization stages of oxygen and carbon (i.e., \ion{O}{vii}, \ion{O}{viii}, and \ion{C}{v}), for which we do not include atomic models. These ionization stages are treated as auxiliary levels, to ensure that the populations of the levels that are modeled are accurate. While it is possible to include these high ionization stages in our models, their level populations are highly dependent on the soft X-ray radiation field, and thus the presence of, e.g., shocks. This should not affect the derived properties of the stars, but does result in a slightly higher uncertainty in the temperature. 

We found that for models with very high temperature ($T_* \ga 190$ kK) the ratios between the \ion{O}{iv} $\lambda$3404-12 \AA, \ion{O}{v} $\lambda$5598 \AA, and \ion{O}{vi} $\lambda$5290 \AA \ lines can no longer be reproduced. For the stars that have such high temperatures we determine the oxygen abundance by adopting models where the oxygen line ratios are closest to the observed values. The uncertainty in the obtained abundance is investigated by fitting each of the individual oxygen lines, while adjusting the other parameters to preserve the overall fit to the spectrum. The range in oxygen abundances that results from this approach is small ($< 5 \% $ changes in the derived mass fractions), and thus we expect them to be accurate. 

Lastly, all spectra show to some extent a very broad emission feature at $\sim$ 21\,000 \AA. This feature is a blend of several emission lines of \ion{C}{iv}, \ion{C}{iii} and \ion{He}{ii}, yet almost none of our models show emission in this region, with the model for LH41-1042 being the exception.

\section{The properties of the single WOs}\label{sec:properties}

In this section we discuss the derived properties of the WO stars and we compare our results to previous research. We then place the WO stars in the Hertzsprung-Russell diagram (HRD) and compare their position with those of the WC stars and with evolutionary tracks. Lastly, we discuss the mass-loss properties of the sample.

\subsection{Comparison with previous results}

Stellar temperatures have been derived for WR102 and WR142 by \cite{sander2012}, who found 200 kK for both stars, in agreement with our results. \cite{crowther2000} report $T_*=150$ kK for BAT99-123 based on their modeling of the far-UV to visible spectrum. {The latter temperature is based on a compromise between the UV and optical diagnostics, as the authors cannot simultaneously fit the \ion{O}{vi} 1032-38 \AA \ resonance line (requiring 120 kK) and \ion{C}{iv} 7700 \AA \ (requiring 170 kK, in agreement with our value). To achieve a fit of both lines likely requires a detailed modeling of the wind acceleration and the clumping properties throughout the outflow.}

For all stars except LH41-1042 one or more measurements of the terminal wind velocity are reported in the literature (see Table~\ref{tab:vinf}). In most cases our values are lower than those previously reported and that are based on full-width at zero line emission and blue edge absorption of ultraviolet P-Cygni profiles. Disparities compared to the first method may be related to a calibration issue. Those with the second method may point to a difference in the treatment of the velocity stratification of the outflow. Several studies report evidence for the presence of two acceleration zones in early WC star winds, in line with theoretical considerations \citep{schmutz1997}. Such a stratification may be modeled using a double-$\beta$ law \citep[e.g., ][]{crowther2000, crowther2002, grafener2005} and can explain the higher $v_{\infty}$ values from P-Cygni profiles that originate further out in the wind than the optical and near-infrared recombination lines used by us.

\begin{table}
\centering
\caption{Comparison of estimates of the terminal wind velocity.}\label{tab:vinf}
\begin{tabular}{l c l }
\hline\hline \\[-8pt]
ID			& $v_{\infty}$	& Reference \\
			&	 {\tiny (km s$^{-1})$}		&	\\
\hline \\[-8pt]	
WR102		&	5000		&	This work			\\
			&	5000		&	\cite{sander2012}	\\
			&	4600		&	\cite{kingsburgh1995a}	\\
WR142		&	4900		&	This work			\\
			&	5000		&	\cite{sander2012}	\\
			&	5500		&	\cite{kingsburgh1995a}	\\
WR93b		&	5000		&	This work			\\
			&	5750		&	\cite{drew2004}	\\
BAT99-123	&	3300		&	This work			\\
			&	4100		&	\cite{crowther2002}	\\
			&	4300		&	\cite{kingsburgh1995a}	\\
DR1			&	2750		&	\cite{tramper2013}	\\
			&	2850		&	\cite{kingsburgh1995b}	\\
\hline
\end{tabular}
\end{table}

Various values of the surface abundances of WO stars have been reported in the literature, an overview of which is presented in Table~\ref{tab:abundance}. Most of these values are derived based on a comparison of equivalent width measurements to recombination line theory, and deviate from our results considerably. This may simply reflect that measuring the equivalent widths of WR lines is difficult due to line blending and the poorly defined continuum. However, the values derived by detailed atmospheric modelling by \cite{crowther2002} do agree with our values within the estimated 10\% uncertainty.

\begin{table}
\centering
\caption{Comparison of estimates of the surface abundances.}\label{tab:abundance}
\begin{tabular}{l l l l }
\hline\hline \\[-8pt]
ID			& $N_{\mathrm{C}}/N_{\mathrm{He}}$	& $N_{\mathrm{O}}/N_{\mathrm{He}}$	& Reference \\
			&			&	\\
\hline \\[-8pt]	
WR102		& 1.50	& 0.45	& This work \\
			& 0.51	& 0.11	& \cite{kingsburgh1995a} \\
			& 0.44\tablefootmark{a}	& 0.25\tablefootmark{a}	& \cite{sander2012} \\
WR142		& 1.00	& 0.16	& This work \\
			& 0.52	& 0.10	& \cite{kingsburgh1995a} \\
			& 0.44\tablefootmark{a}	& 0.25\tablefootmark{a}	& \cite{sander2012} \\
WR93b		& 0.60	& 0.15	& This work \\
			& 0.95	& 0.13	& \cite{drew2004} \\
BAT99-123	& 0.63	& 0.13 	& This work \\
			& 0.51	& 0.11	& \cite{kingsburgh1995a} \\
			& 0.70	& 0.15	& \cite{crowther2002} \\
DR1			& 0.35	& 0.06	& \cite{tramper2013} \\
			& 0.63	& 0.27	& \cite{kingsburgh1995b} \\
\hline
\end{tabular}
\tablefoot{
\tablefoottext{a}{Fixed value (not fitted).}}
\end{table}

\subsection{Hertzsprung-Russell diagram}

\begin{figure}
   	\resizebox{\hsize}{!}{\includegraphics{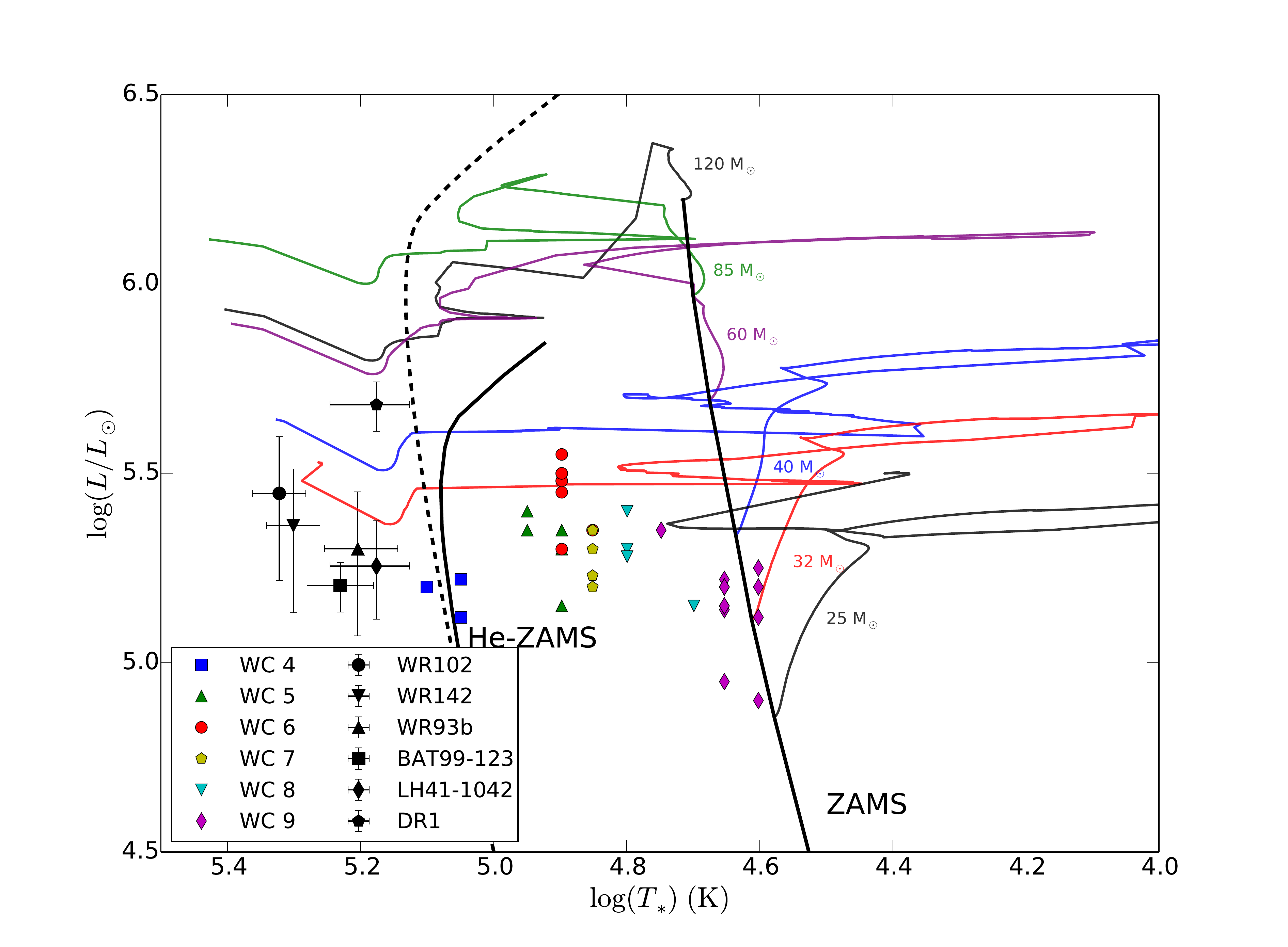}}
  	\caption{Location of the single WO stars in the Hertzsprung-Russell diagram. Also indicated are the WC stars analyzed by \cite{sander2012} and evolutionary tracks for solar metallicity and an initial rotational veloctity of 40\% critical from \cite{ekstrom2012}. The helium zero-age main sequence (He-ZAMS) from {\sc BEC} models (which include envelope inflation) is indicated for solar metallicity (solid line) and SMC metallicity (dashed line). }
	\label{fig:hrd}
\end{figure}

Helium-burning stars that show the products of helium burning in their spectra are expected to be located very close to the helium zero-age main sequence (He-ZAMS) for most of the helium-burning lifetime. They only evolve toward hotter regions after exhausting the helium in their core (see, e.g., Figure~\ref{fig:He_HRD}). 

Figure~\ref{fig:hrd} shows the HRD with our results for the WO stars and the results from \cite{sander2012} for Galactic WC stars. All WO stars are located on the hot side of the He-ZAMS, which is a first indication that they may be post-helium burning objects. The WC stars are located between the ZAMS and He-ZAMS, and thus appear to be too cold for their core-helium burning state. In the past, this discrepancy has often been attributed to the extended photospheres of WR stars. However, the problem remains in recent studies that fully take this effect into account \citep[e.g., ][]{sander2012, hainich2014}. 

A possible solution to explain the positions of WC stars in the HRD is stellar envelope inflation near the Eddington limit. In this context, inflation refers to the extended low-density envelopes in stellar models that reach the Eddington luminosity in their outer layers. This effect is predicted to be strongest at high metallicities (compare the solar and SMC metallicity He-ZAMS in Figure~\ref{fig:hrd}). \cite{grafener2012} could bring the predicted WR radii in agreement with the observations if the inflated sub-surface layers are clumped. \cite{grafener2013} discussed the solution topology of inflated envelopes with optically thick winds and concluded that there are two types of possible solutions: cool stars with clumped inflated envelopes, and hot stars without envelope inflation but a lower mass-loss rate. While the former complies with the properties of late-type WC stars, the latter may represent the small groups of WO stars and early-type WC stars.

\begin{figure}
   	\resizebox{\hsize}{!}{\includegraphics{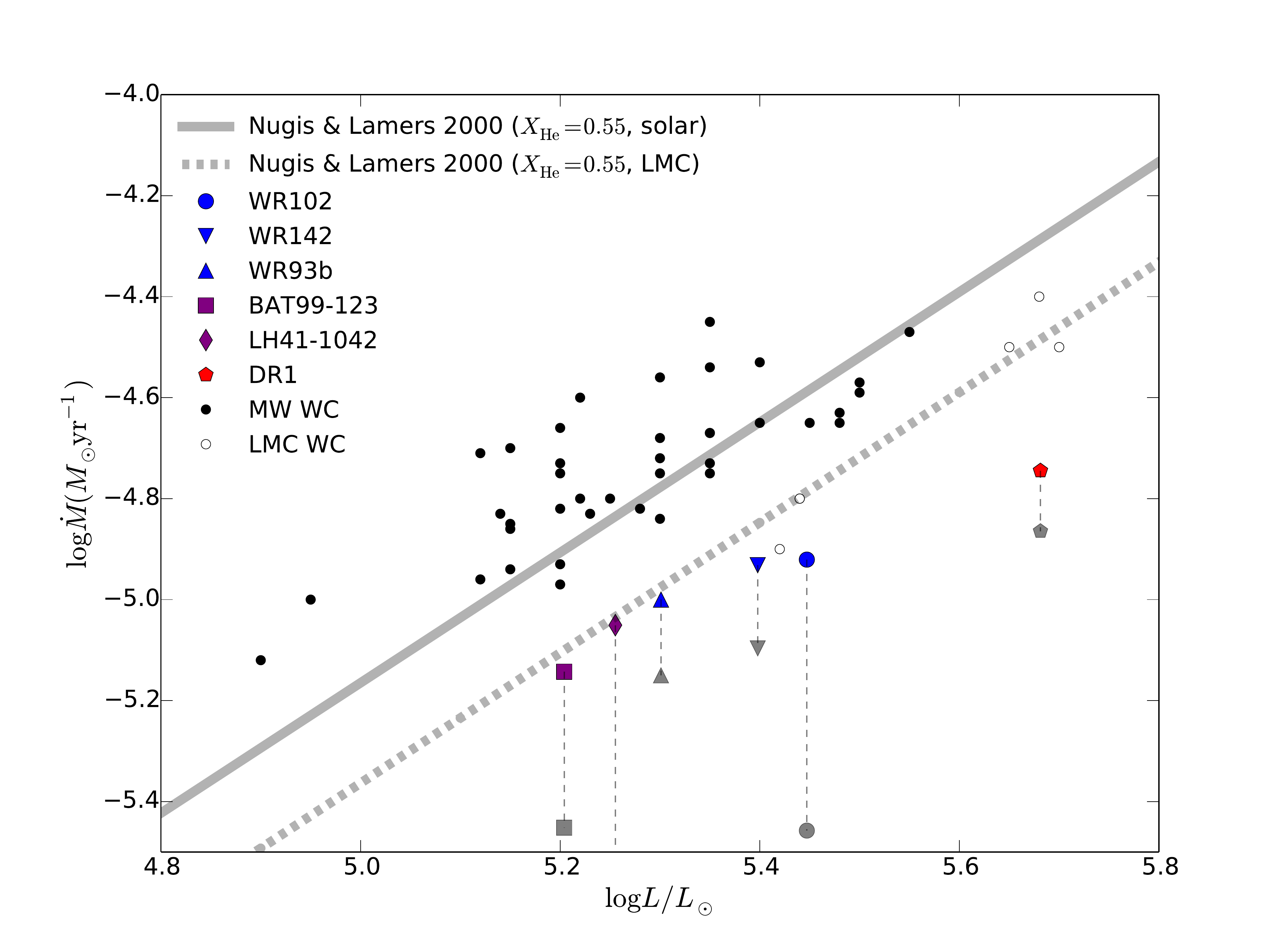}}
  	\caption{Mass loss versus luminosity relation with the location of the WO stars. Results for Galactic WC stars from \cite{sander2012} and LMC WC stars from \cite{crowther2002} are also indicated. Also plotted are the prediction for WC abundances of \cite{nugis2000} with an additional $Z_{\mathrm{Fe}}^{0.66}$ metallicity dependence \citep{vink2005}. The grey symbols indicate the \cite{nugis2000} predictions for the WO stars, which underpredict the observed mass-loss rates.}
	\label{fig:mdotL_nugis}
\end{figure}

The evolutionary tracks for solar metallicity from \cite{ekstrom2012} are also plotted in Figure~\ref{fig:hrd}. The tracks do not reach the region of the HRD where most of the WO stars are located, which indicates that single-star models cannot currently explain them. However, if the mass loss of stars with masses less than $\sim 32 M_{\odot}$ is higher than used in the tracks, these stars might reach the WO star region. In this case, the WO stars seem to be the descendants of stars with initial masses in the range $25 M_{\odot} \la M_{\mathrm{ini}} \la 60 M_{\odot}$. The location of the He-ZAMS (for $Z=Z_{\odot}$ and $Z=0.2Z_{\odot}$) that is plotted in Figure~\ref{fig:hrd} corresponds to the helium-burning models used in Section~\ref{sec:lifetime}. Towards higher luminosities, the He-ZAMS bends towards lower temperatures which is caused by inflation. Again, the location of the WO stars favors them to be post-core helium burning, regardless of any uncertainties in the metallicity.

The evolutionary state of the WO stars is further discussed in Sections~\ref{sec:lifetime} and \ref{sec:nature}.

\subsection{Mass-loss properties}

Figure~\ref{fig:mdotL_nugis} compares the mass-loss rates found for the WO stars to those of the WC stars from \citet[][MW stars]{sander2012} and \citet[][LMC stars]{crowther2002}. Also plotted are various forms of the empirical mass-loss predictions from \cite{nugis2000} for hydrogen-free WR stars:
\begin{equation}
	\log{\dot{M}} = -11 + 1.29 \log{\frac{L}{L_{\odot}}} + 1.7 \log{X_{\mathrm{He}}} + 0.5 \log{Z}.\label{eq:mdotL_nugis}
\end{equation}
\noindent Here, $X_{\mathrm{He}}$ is the helium mass-fraction, and $Z = 1 - X_{\mathrm{He}}$. $Z$ is thus almost equal to the sum of the carbon and oxygen mass fractions. \cite{sander2012} find that the mass-loss rates for their WC stars are compatible with these predictions when using their average carbon and oxygen mass-fractions ($X_{\mathrm{He}} = 0.55, Z = 0.45$, upper grey line in Figure~\ref{fig:mdotL_nugis}). \cite{crowther2002} find an offset of approximately $-0.2$ dex for the mass-loss rates of LMC WC stars, corresponding to a metallicity dependence of $\sim Z_{\mathrm{Fe}}^{0.5}$. \cite{vink2005} predict a scaling of the mass-loss rates with $Z_{\mathrm{Fe}}^{0.66}$ for WC stars with metallicities $0.1 \la Z_{\mathrm{Fe}}/Z_{\mathrm{Fe, \odot}} \la 1$, which we implement in Figure~\ref{fig:mdotL_nugis}. 

The \cite{nugis2000} rates with the \cite{vink2005} metallicity scaling match the observed WC mass-loss rates from \cite{sander2012} and \cite{crowther2002} well. However, they severely underpredict the mass loss of the WO stars, which have a very low helium abundance. This is most notable for WR102, for which the predicted mass-loss rate is about 0.5 dex lower than the observed rate. This needs to be taken into account when comparing to evolutionary predictions that use the \cite{nugis2000} mass-loss prescription.

\begin{figure}
  	\resizebox{\hsize}{!}{\includegraphics{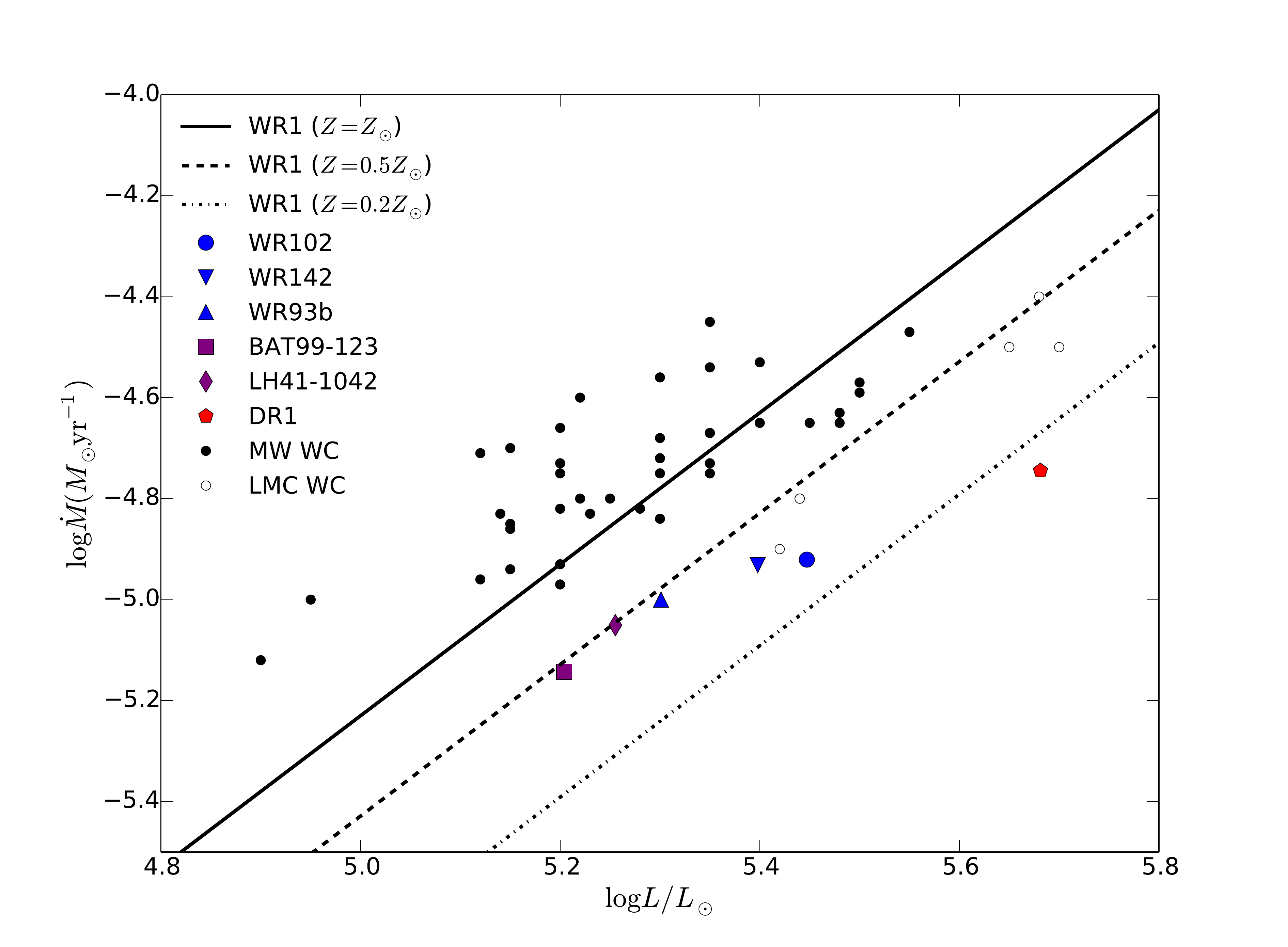}}
	\caption{Same as Figure~\ref{fig:mdotL_nugis}, but with the WR1 relation of \cite{yoon2005} adjusted to have a metallicity dependence of  $Z_{\mathrm{Fe}}^{0.66}$ \citep{vink2005}.}
	\label{fig:mdotL}
\end{figure}

\begin{figure}
   	\resizebox{\hsize}{!}{\includegraphics{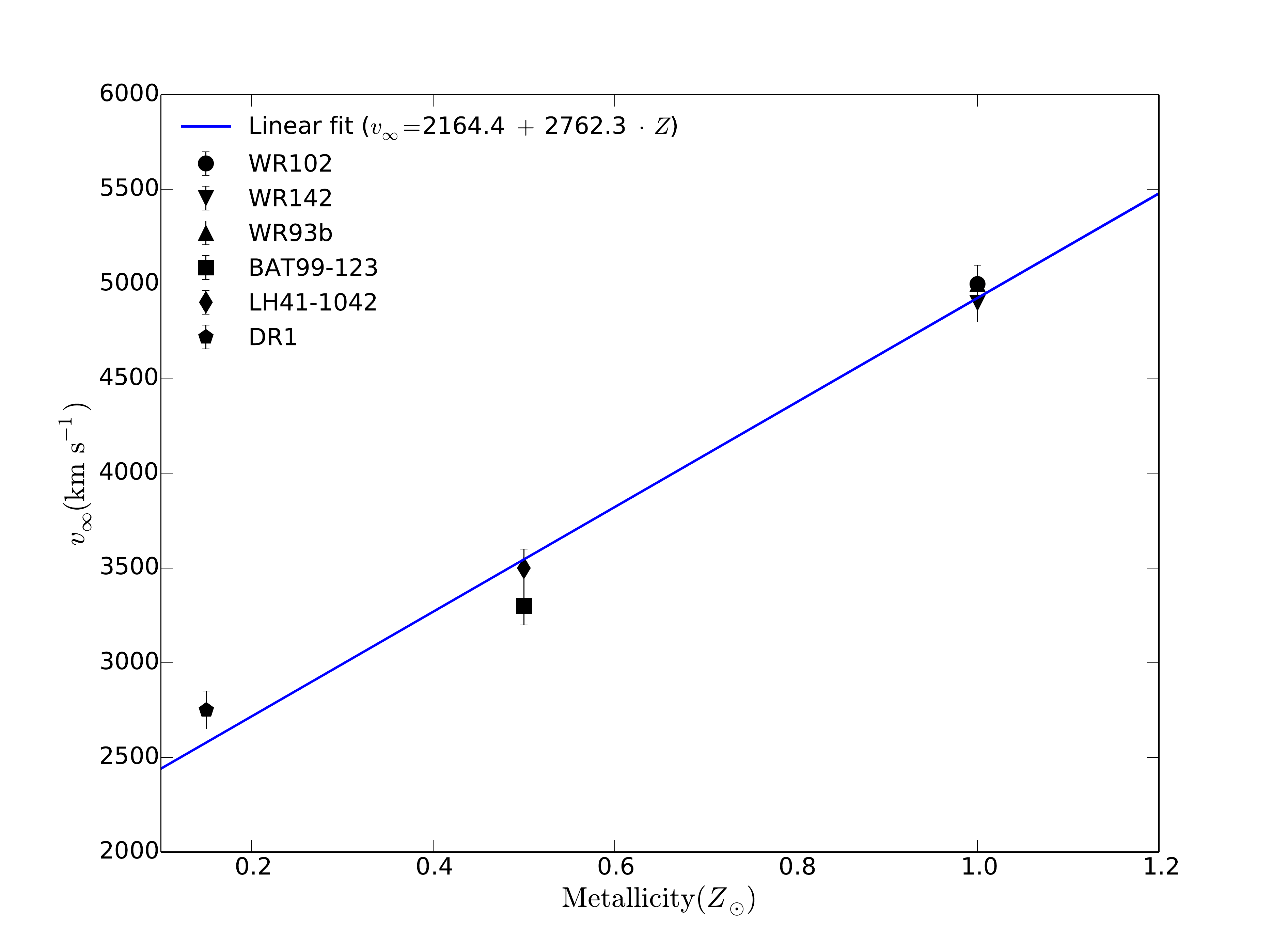}}
  	\caption{Scaling of the terminal wind velocity in WO stars with metallicity.}
	\label{fig:vinfZ}
\end{figure}

Figure~\ref{fig:mdotL} shows the same comparison as Figure~\ref{fig:mdotL_nugis}, but with the mass-loss relation WR1 from \cite{yoon2005}, which is a scaled down version of the results of \cite{hamann1995}. We adjust this relation to have the same metallicity dependence as applied in Figure~\ref{fig:mdotL_nugis}, yielding:
\begin{equation}
 	\log{\dot{M}} = -12.73 + 1.5 \log{\frac{L}{L_{\odot}}} + 0.66 \log{\frac{Z_{\mathrm{Fe}}}{Z_{\mathrm{Fe}, \odot}}}. \label{eq:mdotL}
\end{equation}
\noindent Again, the mass-loss rates of the WC stars are well represented by this relation. The LMC and IC1613 WO stars also have a mass loss that is close to the predicted value. Only the mass-loss rate of the galactic WO stars is not well reproduced by Equation~\ref{eq:mdotL}.

In line with earlier studies \citep[e.g., ][]{crowther2006}, the terminal wind velocities that we derive scale with metalliciy. This behavior is shown in Figure~\ref{fig:vinfZ}. The metallicity dependence can be well represented by a linear relation.

\section{Evolutionary state: remaining lifetime and final fate}\label{sec:lifetime}

\begin{figure}
   	\resizebox{\hsize}{!}{\includegraphics{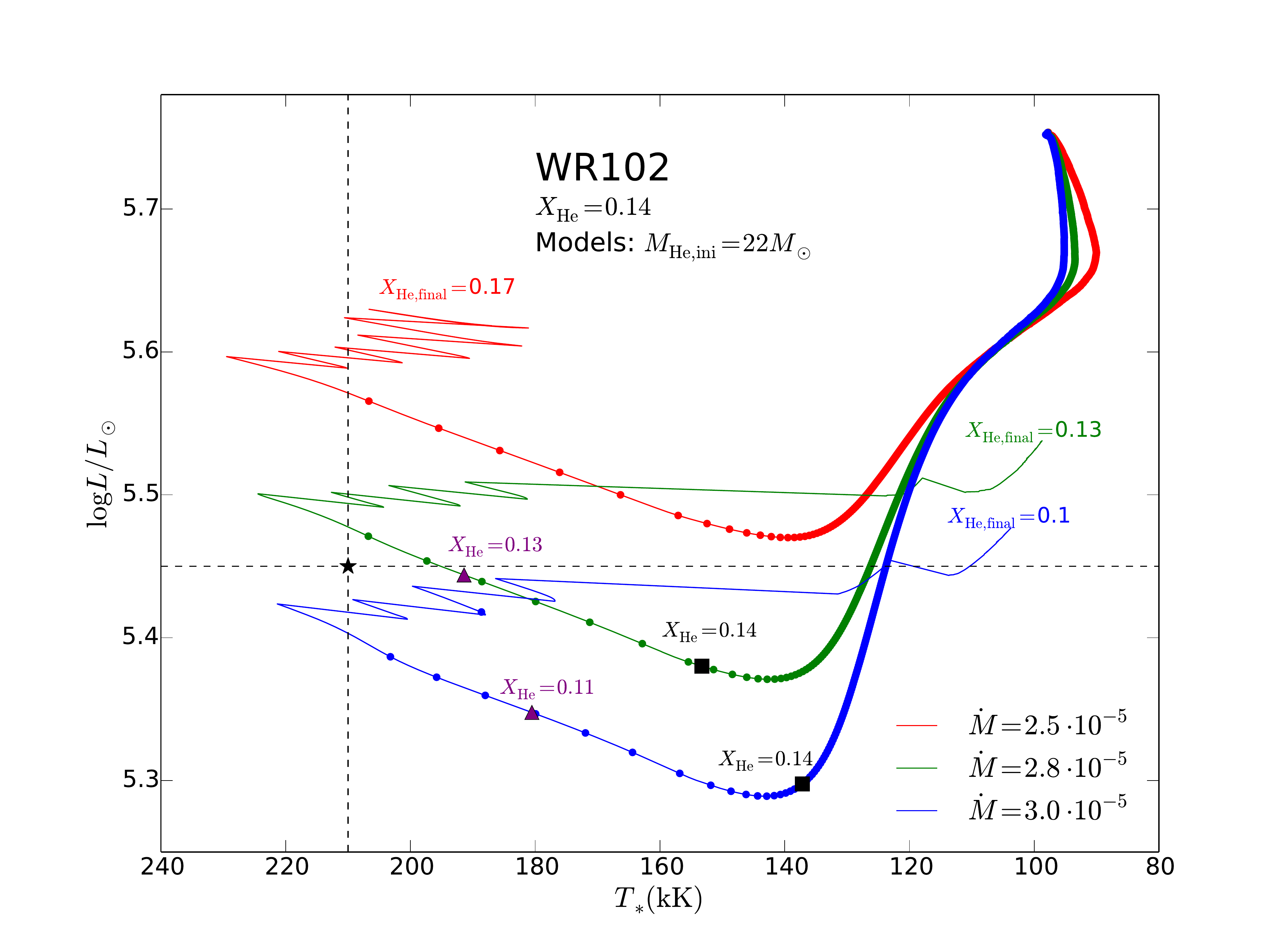}}
  	\caption{Hertzsprung-Russell diagram with evolutionary models for WR102, which start with a helium star. The best-fit parameters for WR102 are indicated. The different tracks correspond to different mass-loss rates. The dots indicate 1000 year time steps.}
	\label{fig:He_HRD}
\end{figure}

The derived properties of the WO stars allow us to estimate the remaining lifetime of these stars, as well as to predict the type of SNe they will produce. As helium burns at a more or less constant rate, the currently observed surface composition corresponds to material that was in the (fully mixed) convective core of the star at a time that is roughly proportional to $(1-X_{\mathrm{He}}) \times \tau_{\mathrm{He}}$, with $\tau_{\mathrm{He}}$ the duration of core-He burning. In Paper I, we used this relation to put a limit on the current stage of helium-burning for DR1.

In this work, we take a more detailed approach. We use the Binary Evolution Code \citep[{\sc BEC}, ][]{yoon2006, yoon2010, brott2011} to model the WR stars as non-rotating hydrogen-free helium stars. {\sc BEC} is a state-of-the-art, one-dimensional hydrodynamic implicit Lagrangian code, and is well-suited to investigate stars that evolve close to the Eddingon limit \citep{kohler2015}.

In particular, convection is treated within the framework of the standard Mixing Length Theory \citep{bohm-vitense1958} with the mixing length set to 1.5 times the local pressure scale-height. All the models were set to hydrostatic equilibrium. The models are evolved from the He-ZAMS until at least core-oxygen ignition.

We calculate evolutionary models for each of the WO stars, with the aim to reproduce the stellar temperature and luminosity at the point where the surface helium abundance of the models equals the observed values (see Table~\ref{tab:abundances}). We use the metallicities listed in Table~\ref{tab:stars}. As shown before, the mass-loss rates from \cite{nugis2000} that are normally used in these evolutionary models are not representative for our stars. We mitigate this by adopting a constant mass-loss rate throughout the evolution. We explore combinations of the mass-loss rate and initial helium star mass to reproduce the observed surface helium abundance, temperature and luminosity. 

Figure~\ref{fig:He_HRD} shows the stellar evolutionary tracks for WR102 in the HRD, computed with an initial helium mass of $M_{\mathrm{He, ini}} = 22 M_{\odot}$ and three values for the mass-loss rate. These were the best-fit models for the observed helium mass fraction, luminosity and temperature. Other combinations of initial helium-star masses and mass-loss rates were explored but they were not in as good agreement with the observed stellar parameters. The surface mass fractions of helium (normalized to $X_{\mathrm{He}} + X_{\mathrm{C}} + X_{\mathrm{O}} = 1$) during the late evolutionary stages are indicated along the tracks. Models for the other WO stars are shown in Appendix~\ref{helium_appendix}. 

The models spend most of their lifetime close to the He-ZAMS. They lose mass as a consequence of the applied mass-loss rate causing the luminosity to drop during the core-helium burning phase. After core-helium exhaustion, the stars contract on their thermal timescale and become hotter and brighter. The models are hardly inflated at this stage of evolution because the iron opacity peak (at $T\sim200$ kK) which is responsible for inflation is only partially present in the stars. At some point during the overall contraction phase the helium-shell ignites and the tracks eventually turn redwards because of the mirror principle \citep{kippenhahn1990}. When the models become cooler, envelope inflation may again play a role. The models were computed until oxygen burning and the star will not change its position in the HRD significantly after the models stop, as the envelope structure does not change any further until the star explodes as a supernova.

\begin{table*}
\centering
\caption{Surface mass fractions, estimates of helium star mass and remaining lifetime, and constraints on the mass loss parameters.}\label{tab:abundances}
\begin{tabular}{l c c c c c c c c}
\hline\hline \\[-8pt]
ID			& $X_{\mathrm{He}}$	& $X_{\mathrm{C}}$	& $X_{\mathrm{O}}$	 	&  $M_{\mathrm{He, ini}}$	 & $M_{\mathrm{final}}$	&$t_\mathrm{SN}$	&$\dot{M}_{\mathrm{evol}}$ 	& $f_c$ \\
			&					&				&					&		{\tiny ($M_{\odot}$)}		&	{\tiny ($M_{\odot}$)} & {\tiny (yr)}	& {\tiny $(M_{\odot} \mathrm{yr}^{-1})$}	\\
\hline \\[-8pt]
WR102	&		0.14			&		0.62		&		0.24			&		22.0		&	9.8	& $1\,500$	&	$2.8 \times 10^{-5}$		& $>0.4$		\\
WR142		&		0.26			&		0.54		&		0.21			&		17.0		&	8.8	& $2\,000$	&	$1.7\times 10^{-5}$		& $>0.4$		\\
WR93b		&		0.29			&		0.53		&		0.18			&		17.0		&	8.8	& $8\,000$	&	$1.7\times 10^{-5}$		& $>0.3 $			\\
BAT99-123	&		0.30			&		0.55		&		0.15			&		15.0		&	7.7	&$7\,000$		&	$1.4 \times 10^{-5}$		& $>0.3$		\\
LH41-1042	&		0.22			&		0.60		&		0.18			&		17.0		&	8.4	& $9\,000$	&	$1.8\times 10^{-5}$		& $> 0.4$		\\
DR1		&		0.44			&		0.46		&		0.10			&		23.0		&	15.4	& $17\,000$		&	$1.8 \times 10^{-5}$	& $0.1$ \\
\hline
\end{tabular}
\end{table*}

The stellar tracks in Figure~\ref{fig:He_HRD} that reproduce the observed stellar parameters do so after the models have exhausted the helium in their core \citep[see also ][]{langer1988, langer1989}. In other words, according to our models WR102 is a post-core helium burning star and has a remaining lifetime of less than 2000 years. Furthermore, tracks with mass-loss rates lower than $\sim 2.7 \times 10^{-5} M_{\odot} \mathrm{yr}^{-1}$ never reveal the layer that corresponds to the observed helium mass-fraction. Combined with the determined value of $\dot{M}/\sqrt{f_{\mathrm{c}}}$, this constrains the volume filling factor to $f_{\mathrm{c}}\geq 0.4$.

As the migration to higher temperatures after core-helium exhaustion occurs on a very short timescale, the surface abundances do not change significantly during this period. This implies that a few WC stars with similar surface abundances may be expected to exist. \cite{koesterke1995} indeed find carbon surface mass fractions for WC stars in the range $0.2-0.6$, i.e. covering the $0.4-0.6$ range that we find for the WO stars (Table~\ref{tab:abundances}). The oxygen abundance is not well determined for WC stars \citep[e.g., ][]{crowther2007} so a direct comparison with our results cannot be made.

Table~\ref{tab:abundances} gives the initial helium star mass, predicted remaining lifetime, and constraints on the clumping factor based on the calculated evolutionary models for all the stars. According to our models, all WO stars except maybe DR1 (which is close to the helium terminal-age main sequence) are in the core-contraction phase after core-helium burning. They are likely to explode in less than $10^4$ years. The initial helium star masses are in the range $M_{\mathrm{He, ini}} \sim 15-25 \, M_{\odot}$. This corresponds to the helium-core masses predicted for stars with an initial mass of $M_{\mathrm{ini}} \sim 40-60 \, M_{\odot}$ \citep{ekstrom2012}. However, only the models with very rapid rotation (40\% of critical, Figure~\ref{fig:hrd}) reach the He-ZAMS and hotter regions, while non-rotating models are still covered by a hydrogen-rich envelope. As most massive stars are only modestly rotating \citep[e.g., ][]{ramirez2013}, this may point to a much higher mass loss (either through the stellar wind or binary interactions) in stages prior to the WO phase.

\cite{groh2014} modelled the spectrum of a non-rotating star with an initial mass of $60 M_{\odot}$ during various stages of its evolution, based on the parameters predicted by the \cite{ekstrom2012} evolutionary tracks. These tracks produce a WO spectrum during a very short stage prior to core-helium exhaustion. After core-helium exhaustion the tracks again enter a longer WO phase that lasts until the star explodes. We find that the WO star with an estimated initial mass close to 60 $M_{\odot}$ (WR102) has a luminosity that is much lower then those predicted by the \cite{ekstrom2012} tracks (see Figure~\ref{fig:hrd}). This indicates that more mass has been lost during its evolution compared to the theoretical predictions.

While all observed WO stars are post-core helium burning, but this does not exclude that some of the WC stars are in this phase as well, and thus could explode without ever becoming a WO star. Whether the whole star contracts after core-helium exhaustion (i.e. becomes a WO star) is dependent on the mass loss prior to, and during, the helium-burning phase and must be much higher then the currently used \cite{nugis2000} predictions.

When they end their lives, most of the WO stars have predicted masses below $10 \, M_{\odot}$. The supernovae will almost certainly be of type Ic, as the current fractions of helium at the surface are already too low to produce type Ib SNe \citep{dessart2011}. Although \cite{sander2012} report on a very high rotational velocity for WR102 and WR142, we do not find any indication of rapid rotation for the WO stars. We can therefore not conclude on the possibility of the production of long-duration GRBs during core-collapse.

\subsection{Progenitors of WO stars}

As shown above, the likely progenitors of WO stars have an initial mass of $M_{\mathrm{ini}} = 40-60 M_{\odot}$. This is in very good agreement with the current view of massive star evolution, for instance as summarized by \cite{langer2012}. The evolutionary sequences proposed by \cite{conti1975} and later refined by \cite{langer1994} and \cite{sander2012} also predict WO stars to be the descendants of stars with $M_{\mathrm{ini}} = 45-60 M_{\odot}$. 

As mentioned before, single-star evolutionary tracks that start from the ZAMS cannot reproduce most of the WO stars. However, the evolutionary path of massive stars and the mass-loss rates at each of the various stages of evolution are highly uncertain. In particular, whether a star undergoes an LBV phase, and how much mass is lost during that phase, is currently poorly understood. The question is therefore how the naked helium stars that are needed to explain the WO phenomena are formed. While some WO stars may indeed originate from single-star evolution, binary interactions may also play a role \citep{sana2012}.

If the WO star progenitor is initially the most massive star in a binary system, it might strip its envelope by transferring mass to a massive companion \citep[e.g., ][]{paczynski1971}. The hydrogen-rich material would rejuvenate the companion star, making it appear younger \citep[e.g., ][]{hellings1983, dray2007,de-mink2014}. This scenario might be appropriate for the two WO stars that are in a short-period binary with an O-star companion, WR30a and Sk188. The spectral signature of their companions are clearly visible in the observed spectra (see Figure~\ref{fig:atlas_UVBVIS}). A quantitative analysis of these stars in a later work may reveal the potential role of binarity in the evolution of WO star progenitors. The single WO stars that are discussed in this paper show no spectral signature of a massive companion. However, we cannot exclude the presence of a fainter intermediate-mass companion as in the WO stars might have lost their hydrogen envelope in a common envelope interaction. In that case the companion star is expected to still orbit the WO star but would be too faint to be seen in the spectrum. If the WO stars do instead result from single-star evolution, an LBV phase after core-hydrogen exhaustion may be the likely explanation to reveal the core at the onset of helium burning \citep[as is the case in ][]{groh2014}.

\subsection{The $^{12}\mathrm{C}(\alpha,\gamma)^{16}\mathrm{O}$ thermonuclear cross-section}

\begin{figure}
  	\resizebox{\hsize}{!}{\includegraphics{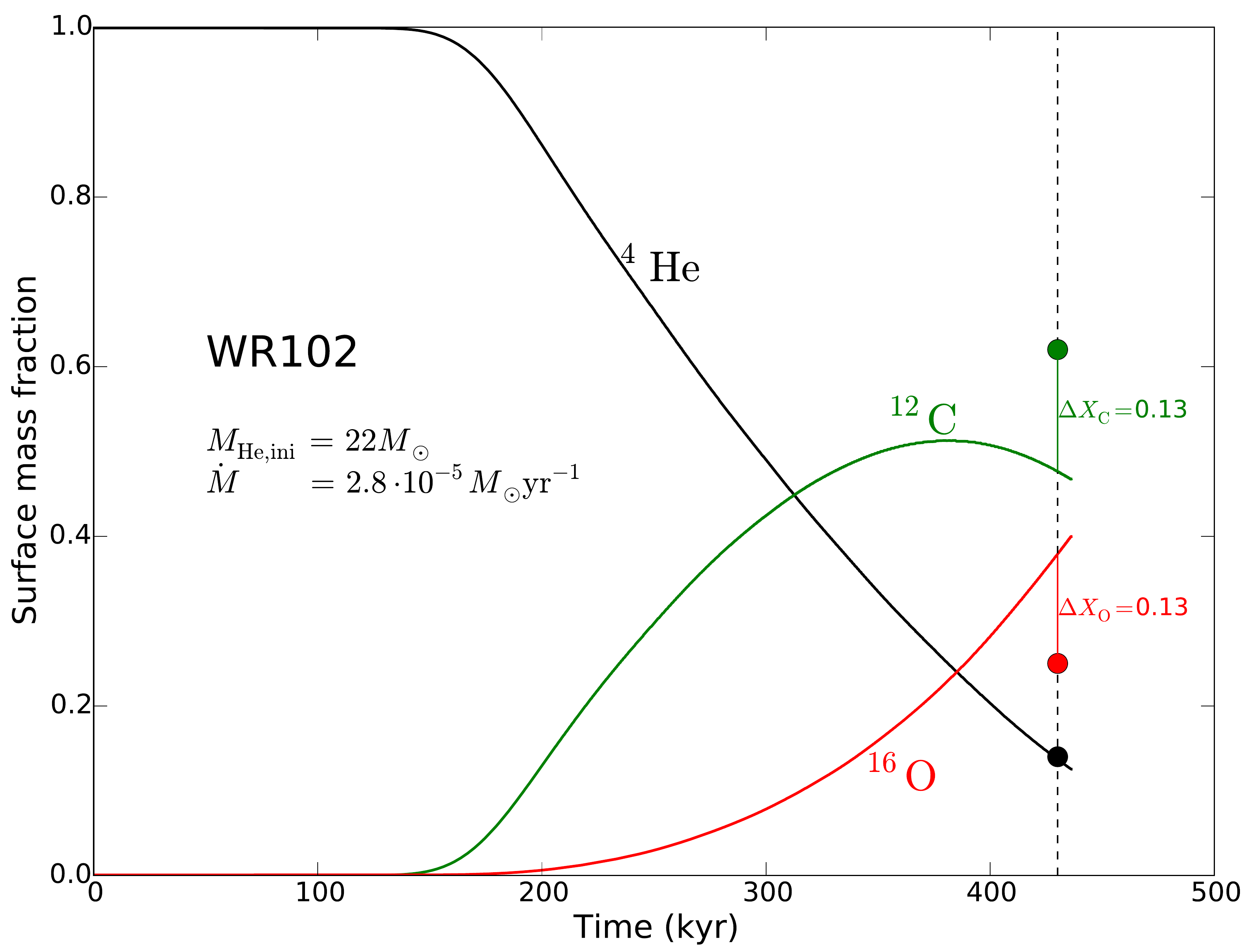}}
 	\caption{Evolution of the surface mass fractions in the helium star model of WR102 since the onset of core-helium burning. The observed mass fractions of helium, carbon, and oxygen are indicated. The model over predicts the $X_{\mathrm{O}}/X_{\mathrm{C}}$ ratio.}
	\label{fig:He_abundances}
\end{figure}

Apart from constraining the lifetime of the WO stars, the observed surface abundances can be used to constrain the elusive $^{12}\mathrm{C}(\alpha,\gamma)^{16}\mathrm{O}$ thermonuclear reaction rate \citep{grafener1998}. The currently used value \citep[0.632 times that of][]{caughlan1985} is based on the solar abundance pattern between oxygen and calcium, but has an uncertainty of about 30\% \citep{weaver1993}. Deviations of this rate strongly influence the supernova yields as well as the pre-supernova evolution \citep{tur2007}.
With their surface abundances corresponding to the core abundances far into the helium-burning stage, WO stars offer a unique opportunity to provide direct constraints on the nuclear reaction rate. Figure~\ref{fig:He_abundances} shows the helium, carbon, and oxygen surface mass fractions as a function of time for the helium star model of WR102. When compared to the observed mass fractions, it is clear that the model overpredicts the oxygen-to-carbon ratio. The same effect is seen in the models for all the other stars. This indicates that the actual $^{12}\mathrm{C}(\alpha,\gamma)^{16}\mathrm{O}$ reaction rate must be lower than the value that is currently being used. Placing more firm constraints on this reaction rate will be the focus of a separate study.

\section{Summary and conclusions}\label{sec:nature}

In this paper we have presented a detailed spectroscopic analysis of single WO stars. We have constrained the evolutionary status of the sources using tailored evolutionary models.  For the stellar properties we find that:

\begin{itemize}
\item WO stars are extremely hot, with temperatures ranging from 150 kK up to 210 kK.  In the Hertzsprung-Russell diagram 
          they  are located at the hot side of the helium zero-age main sequence;
\item	 The luminosities range from $1.2\times10^5 \, L_{\odot}$ to $3.2 \times 10^5 \, L_{\odot}$, comparable to the luminosities of WC stars. Only DR1 is much brighter than WC stars, with a luminosity of $5\times 10^5 \, L_{\odot}$.
\item The helium surface mass fraction is typically 20-30\%, but ranges from 44\% for the coolest star to as low as 14\% for 
          the hottest star. The oxygen mass fractions reach values up to 25\%;
\item The stars lose mass at a rate of $0.8-1.5 \times 10^{-5} M_{\odot} \, \mathrm{yr}^{-1}$, which is higher by a factor of about 2-3 than suggested by
          the predictions by \cite{nugis2000};
\item The surface abundances of WO stars correspond to material that was formed very late into the core-helium burning stage, 
          and the stars must have experienced severe mass loss during this stage to reveal the observed layers. This translates to strict constraints on the volume filling factor of the wind medium, requiring $f_\mathrm{c} > 0.3$ for most 
          of the stars.
\end{itemize}

The derived properties of the WO stars can be reproduced for helium stars with $M_{\mathrm{He, ini}} \sim 15-25 \, M_{\odot}$, and
suggest initial masses of $M_{\mathrm{ini}} \sim 40-60 \, M_{\odot}$.  Their extremely high temperatures are consistent with the contraction of the 
star after it exhausts its helium in the core. Together with the low surface helium abundances, this firmly establishes that WO stars
are post-core helium burning massive stars. They will likely
explode as type Ic supernovae in $10^3-10^4$ years.  Thus, the WO class indeed represents a short, final stage in the 
evolution of massive stars.

\begin{acknowledgements}
We thank the anonymous referee for his thorough reading of the manuscript and his suggestions to improve the paper.
We thank Bram Ochsendorf and Bertrand Lemasle for executing part of the observations.
This research has made use of the SIMBAD database, operated at CDS, Strasbourg, France
\end{acknowledgements}

\bibliographystyle{aa}
\bibliography{WOarchiv}

\begin{thebibliography}{64}
\expandafter\ifx\csname natexlab\endcsname\relax\def\natexlab#1{#1}\fi

\bibitem[{{Barlow} \& {Hummer}(1982)}]{barlow1982}
{Barlow}, M.~J. \& {Hummer}, D.~G. 1982, in IAU Symposium, Vol.~99, Wolf-Rayet
  Stars: Observations, Physics, Evolution, ed. C.~W.~H. {De Loore} \& A.~J.
  {Willis}, 387--392

\bibitem[{{B{\"o}hm-Vitense}(1958)}]{bohm-vitense1958}
{B{\"o}hm-Vitense}, E. 1958, \zap, 46, 108

\bibitem[{{Brott} {et~al.}(2011){Brott}, {de Mink}, {Cantiello}, {Langer}, {de
  Koter}, {Evans}, {Hunter}, {Trundle}, \& {Vink}}]{brott2011}
{Brott}, I., {de Mink}, S.~E., {Cantiello}, M., {et~al.} 2011, \aap, 530, A115

\bibitem[{{Cardelli} {et~al.}(1989){Cardelli}, {Clayton}, \&
  {Mathis}}]{cardelli1989}
{Cardelli}, J.~A., {Clayton}, G.~C., \& {Mathis}, J.~S. 1989, \apj, 345, 245

\bibitem[{{Caughlan} {et~al.}(1985){Caughlan}, {Fowler}, {Harris}, \&
  {Zimmerman}}]{caughlan1985}
{Caughlan}, G.~R., {Fowler}, W.~A., {Harris}, M.~J., \& {Zimmerman}, B.~A.
  1985, Atomic Data and Nuclear Data Tables, 32, 197

\bibitem[{{Conti}(1975)}]{conti1975}
{Conti}, P.~S. 1975, Memoires of the Societe Royale des Sciences de Liege, 9,
  193

\bibitem[{{Crowther}(2007)}]{crowther2007}
{Crowther}, P.~A. 2007, \araa, 45, 177

\bibitem[{{Crowther} {et~al.}(1998){Crowther}, {De Marco}, \&
  {Barlow}}]{crowther1998}
{Crowther}, P.~A., {De Marco}, O., \& {Barlow}, M.~J. 1998, \mnras, 296, 367

\bibitem[{{Crowther} {et~al.}(2002){Crowther}, {Dessart}, {Hillier}, {Abbott},
  \& {Fullerton}}]{crowther2002}
{Crowther}, P.~A., {Dessart}, L., {Hillier}, D.~J., {Abbott}, J.~B., \&
  {Fullerton}, A.~W. 2002, \aap, 392, 653

\bibitem[{{Crowther} {et~al.}(2000){Crowther}, {Fullerton}, {Hillier},
  {Brownsberger}, {Dessart}, {Willis}, {De Marco}, {Barlow}, {Hutchings},
  {Massa}, {Morton}, \& {Sonneborn}}]{crowther2000}
{Crowther}, P.~A., {Fullerton}, A.~W., {Hillier}, D.~J., {et~al.} 2000, \apjl,
  538, L51

\bibitem[{{Crowther} \& {Hadfield}(2006)}]{crowther2006}
{Crowther}, P.~A. \& {Hadfield}, L.~J. 2006, \aap, 449, 711

\bibitem[{{de Koter} {et~al.}(1997){de Koter}, {Heap}, \&
  {Hubeny}}]{de-koter1997}
{de Koter}, A., {Heap}, S.~R., \& {Hubeny}, I. 1997, \apj, 477, 792

\bibitem[{{de Mink} {et~al.}(2014){de Mink}, {Sana}, {Langer}, {Izzard}, \&
  {Schneider}}]{de-mink2014}
{de Mink}, S.~E., {Sana}, H., {Langer}, N., {Izzard}, R.~G., \& {Schneider},
  F.~R.~N. 2014, \apj, 782, 7

\bibitem[{{Dessart} {et~al.}(2011){Dessart}, {Hillier}, {Livne}, {Yoon},
  {Woosley}, {Waldman}, \& {Langer}}]{dessart2011}
{Dessart}, L., {Hillier}, D.~J., {Livne}, E., {et~al.} 2011, \mnras, 414, 2985

\bibitem[{{Dray} \& {Tout}(2007)}]{dray2007}
{Dray}, L.~M. \& {Tout}, C.~A. 2007, \mnras, 376, 61

\bibitem[{{Drew} {et~al.}(2004){Drew}, {Barlow}, {Unruh}, {Parker}, {Wesson},
  {Pierce}, {Masheder}, \& {Phillipps}}]{drew2004}
{Drew}, J.~E., {Barlow}, M.~J., {Unruh}, Y.~C., {et~al.} 2004, \mnras, 351, 206

\bibitem[{{Ekstr{\"o}m} {et~al.}(2012){Ekstr{\"o}m}, {Georgy}, {Eggenberger},
  {Meynet}, {Mowlavi}, {Wyttenbach}, {Granada}, {Decressin}, {Hirschi},
  {Frischknecht}, {Charbonnel}, \& {Maeder}}]{ekstrom2012}
{Ekstr{\"o}m}, S., {Georgy}, C., {Eggenberger}, P., {et~al.} 2012, \aap, 537,
  A146

\bibitem[{{Gibson}(2000)}]{gibson2000}
{Gibson}, B.~K. 2000, \memsai, 71, 693

\bibitem[{{Gr{\"a}fener} \& {Hamann}(2005)}]{grafener2005}
{Gr{\"a}fener}, G. \& {Hamann}, W.-R. 2005, \aap, 432, 633

\bibitem[{{Gr\"afener} {et~al.}(1998){Gr\"afener}, {Hamann}, {Hillier}, \&
  {Koesterke}}]{grafener1998}
{Gr\"afener}, G., {Hamann}, W.-R., {Hillier}, D.~J., \& {Koesterke}, L. 1998,
  \aap, 329, 190

\bibitem[{{Gr{\"a}fener} {et~al.}(2012){Gr{\"a}fener}, {Owocki}, \&
  {Vink}}]{grafener2012}
{Gr{\"a}fener}, G., {Owocki}, S.~P., \& {Vink}, J.~S. 2012, \aap, 538, A40

\bibitem[{{Gr{\"a}fener} \& {Vink}(2013)}]{grafener2013}
{Gr{\"a}fener}, G. \& {Vink}, J.~S. 2013, \aap, 560, A6

\bibitem[{{Groh} {et~al.}(2014){Groh}, {Meynet}, {Ekstr{\"o}m}, \&
  {Georgy}}]{groh2014}
{Groh}, J.~H., {Meynet}, G., {Ekstr{\"o}m}, S., \& {Georgy}, C. 2014, \aap,
  564, A30

\bibitem[{{Hainich} {et~al.}(2014){Hainich}, {R{\"u}hling}, {Todt}, {Oskinova},
  {Liermann}, {Gr{\"a}fener}, {Foellmi}, {Schnurr}, \& {Hamann}}]{hainich2014}
{Hainich}, R., {R{\"u}hling}, U., {Todt}, H., {et~al.} 2014, \aap, 565, A27

\bibitem[{{Hamann} \& {Koesterke}(1998)}]{hamann1998}
{Hamann}, W.-R. \& {Koesterke}, L. 1998, \aap, 335, 1003

\bibitem[{{Hamann} {et~al.}(1995){Hamann}, {Koesterke}, \&
  {Wessolowski}}]{hamann1995}
{Hamann}, W.-R., {Koesterke}, L., \& {Wessolowski}, U. 1995, \aap, 299, 151

\bibitem[{{Hellings}(1983)}]{hellings1983}
{Hellings}, P. 1983, \apss, 96, 37

\bibitem[{{Hillier} \& {Miller}(1998)}]{hillier1998}
{Hillier}, D.~J. \& {Miller}, D.~L. 1998, \apj, 496, 407

\bibitem[{{Hillier} \& {Miller}(1999)}]{hillier1999}
{Hillier}, D.~J. \& {Miller}, D.~L. 1999, \apj, 519, 354

\bibitem[{{Kingsburgh} \& {Barlow}(1995)}]{kingsburgh1995b}
{Kingsburgh}, R.~L. \& {Barlow}, M.~J. 1995, \aap, 295, 171

\bibitem[{{Kingsburgh} {et~al.}(1995){Kingsburgh}, {Barlow}, \&
  {Storey}}]{kingsburgh1995a}
{Kingsburgh}, R.~L., {Barlow}, M.~J., \& {Storey}, P.~J. 1995, \aap, 295, 75

\bibitem[{{Kippenhahn} \& {Weigert}(1990)}]{kippenhahn1990}
{Kippenhahn}, R. \& {Weigert}, A. 1990, {Stellar Structure and Evolution}

\bibitem[{{Koesterke} \& {Hamann}(1995)}]{koesterke1995}
{Koesterke}, L. \& {Hamann}, W.-R. 1995, \aap, 299, 503

\bibitem[{{K{\"o}hler} {et~al.}(2015){K{\"o}hler}, {Langer}, {de Koter}, {de
  Mink}, {Crowther}, {Evans}, {Gr{\"a}fener}, {Sana}, {Sanyal}, {Schneider}, \&
  {Vink}}]{kohler2015}
{K{\"o}hler}, K., {Langer}, N., {de Koter}, A., {et~al.} 2015, \aap, 573, A71

\bibitem[{{Langer}(1989)}]{langer1989}
{Langer}, N. 1989, \aap, 210, 93

\bibitem[{{Langer}(2012)}]{langer2012}
{Langer}, N. 2012, \araa, 50, 107

\bibitem[{{Langer} {et~al.}(1994){Langer}, {Hamann}, {Lennon}, {Najarro},
  {Pauldrach}, \& {Puls}}]{langer1994}
{Langer}, N., {Hamann}, W.-R., {Lennon}, M., {et~al.} 1994, \aap, 290, 819

\bibitem[{{Langer} {et~al.}(1988){Langer}, {Kiriakidis}, {El Eid}, {Fricke}, \&
  {Weiss}}]{langer1988}
{Langer}, N., {Kiriakidis}, M., {El Eid}, M.~F., {Fricke}, K.~J., \& {Weiss},
  A. 1988, \aap, 192, 177

\bibitem[{{Massey} {et~al.}(2014){Massey}, {Neugent}, {Morrell}, \&
  {Hillier}}]{massey2014}
{Massey}, P., {Neugent}, K.~F., {Morrell}, N., \& {Hillier}, D.~J. 2014, \apj,
  788, 83

\bibitem[{{Massey} {et~al.}(2005){Massey}, {Puls}, {Pauldrach}, {Bresolin},
  {Kudritzki}, \& {Simon}}]{massey2005}
{Massey}, P., {Puls}, J., {Pauldrach}, A.~W.~A., {et~al.} 2005, \apj, 627, 477

\bibitem[{{Massey} {et~al.}(2000){Massey}, {Waterhouse}, \&
  {DeGioia-Eastwood}}]{massey2000}
{Massey}, P., {Waterhouse}, E., \& {DeGioia-Eastwood}, K. 2000, \aj, 119, 2214

\bibitem[{{Moffat} \& {Seggewiss}(1984)}]{moffat1984}
{Moffat}, A.~F.~J. \& {Seggewiss}, W. 1984, \aaps, 58, 117

\bibitem[{{Neugent} {et~al.}(2012){Neugent}, {Massey}, \&
  {Morrell}}]{neugent2012}
{Neugent}, K.~F., {Massey}, P., \& {Morrell}, N. 2012, \aj, 144, 162

\bibitem[{{Nugis} \& {Lamers}(2000)}]{nugis2000}
{Nugis}, T. \& {Lamers}, H.~J.~G.~L.~M. 2000, \aap, 360, 227

\bibitem[{{O'Donnell}(1994)}]{odonnell1994}
{O'Donnell}, J.~E. 1994, \apj, 422, 158

\bibitem[{{Paczy{\'n}ski}(1971)}]{paczynski1971}
{Paczy{\'n}ski}, B. 1971, \araa, 9, 183

\bibitem[{{Ram{\'{\i}}rez-Agudelo} {et~al.}(2013){Ram{\'{\i}}rez-Agudelo},
  {Sim{\'o}n-D{\'{\i}}az}, {Sana}, {de Koter}, {Sab{\'{\i}}n-Sanjul{\'{\i}}an},
  {de Mink}, {Dufton}, {Gr{\"a}fener}, {Evans}, {Herrero}, {Langer}, {Lennon},
  {Ma{\'{\i}}z Apell{\'a}niz}, {Markova}, {Najarro}, {Puls}, {Taylor}, \&
  {Vink}}]{ramirez2013}
{Ram{\'{\i}}rez-Agudelo}, O.~H., {Sim{\'o}n-D{\'{\i}}az}, S., {Sana}, H.,
  {et~al.} 2013, \aap, 560, A29

\bibitem[{{Sana} {et~al.}(2012){Sana}, {de Mink}, {de Koter}, {Langer},
  {Evans}, {Gieles}, {Gosset}, {Izzard}, {Le Bouquin}, \&
  {Schneider}}]{sana2012}
{Sana}, H., {de Mink}, S.~E., {de Koter}, A., {et~al.} 2012, Science, 337, 444

\bibitem[{{Sander} {et~al.}(2012){Sander}, {Hamann}, \& {Todt}}]{sander2012}
{Sander}, A., {Hamann}, W.-R., \& {Todt}, H. 2012, \aap, 540, A144

\bibitem[{{Sanduleak}(1971)}]{sanduleak1971}
{Sanduleak}, N. 1971, \apjl, 164, L71

\bibitem[{{Schmutz}(1997)}]{schmutz1997}
{Schmutz}, W. 1997, \aap, 321, 268

\bibitem[{{Schmutz} {et~al.}(1989){Schmutz}, {Hamann}, \&
  {Wessolowski}}]{schmutz1989}
{Schmutz}, W., {Hamann}, W.-R., \& {Wessolowski}, U. 1989, \aap, 210, 236

\bibitem[{{Tramper} {et~al.}(2013){Tramper}, {Gr{\"a}fener}, {Hartoog}, {Sana},
  {de Koter}, {Vink}, {Ellerbroek}, {Langer}, {Garcia}, {Kaper}, \& {de
  Mink}}]{tramper2013}
{Tramper}, F., {Gr{\"a}fener}, G., {Hartoog}, O.~E., {et~al.} 2013, \aap, 559,
  A72

\bibitem[{{Tur} {et~al.}(2007){Tur}, {Heger}, \& {Austin}}]{tur2007}
{Tur}, C., {Heger}, A., \& {Austin}, S.~M. 2007, \apj, 671, 821

\bibitem[{{Tur} {et~al.}(2006){Tur}, {Wuosmaa}, {Austin}, {Lighthall},
  {Marley}, {Goodman}, {Bos}, {Heger}, {Woosley}, \& {Hoffman}}]{tur2006}
{Tur}, C., {Wuosmaa}, A., {Austin}, S.~M., {et~al.} 2006, in APS Division of
  Nuclear Physics Meeting Abstracts, C3

\bibitem[{{Vernet} {et~al.}(2011){Vernet}, {Dekker}, {D'Odorico}, {Kaper},
  {Kjaergaard}, {Hammer}, {Randich}, {Zerbi}, {Groot}, {Hjorth}, {Guinouard},
  {Navarro}, {Adolfse}, {Albers}, {Amans}, {Andersen}, {Andersen}, {Binetruy},
  {Bristow}, {Castillo}, {Chemla}, {Christensen}, {Conconi}, {Conzelmann},
  {Dam}, {de Caprio}, {de Ugarte Postigo}, {Delabre}, {di Marcantonio},
  {Downing}, {Elswijk}, {Finger}, {Fischer}, {Flores}, {Fran{\c c}ois},
  {Goldoni}, {Guglielmi}, {Haigron}, {Hanenburg}, {Hendriks}, {Horrobin},
  {Horville}, {Jessen}, {Kerber}, {Kern}, {Kiekebusch}, {Kleszcz}, {Klougart},
  {Kragt}, {Larsen}, {Lizon}, {Lucuix}, {Mainieri}, {Manuputy}, {Martayan},
  {Mason}, {Mazzoleni}, {Michaelsen}, {Modigliani}, {Moehler}, {M{\o}ller},
  {Norup S{\o}rensen}, {N{\o}rregaard}, {P{\'e}roux}, {Patat}, {Pena}, {Pragt},
  {Reinero}, {Rigal}, {Riva}, {Roelfsema}, {Royer}, {Sacco}, {Santin},
  {Schoenmaker}, {Spano}, {Sweers}, {Ter Horst}, {Tintori}, {Tromp}, {van
  Dael}, {van der Vliet}, {Venema}, {Vidali}, {Vinther}, {Vola}, {Winters},
  {Wistisen}, {Wulterkens}, \& {Zacchei}}]{vernet2011}
{Vernet}, J., {Dekker}, H., {D'Odorico}, S., {et~al.} 2011, \aap, 536, A105

\bibitem[{{Vink} \& {de Koter}(2005)}]{vink2005}
{Vink}, J.~S. \& {de Koter}, A. 2005, \aap, 442, 587

\bibitem[{{Vink} \& {Gr{\"a}fener}(2012)}]{vink2012}
{Vink}, J.~S. \& {Gr{\"a}fener}, G. 2012, \apjl, 751, L34

\bibitem[{{Weaver} \& {Woosley}(1993)}]{weaver1993}
{Weaver}, T.~A. \& {Woosley}, S.~E. 1993, \physrep, 227, 65

\bibitem[{{Woosley} \& {Bloom}(2006)}]{woosley2006}
{Woosley}, S.~E. \& {Bloom}, J.~S. 2006, \araa, 44, 507

\bibitem[{{Yoon} {et~al.}(2012){Yoon}, {Gr{\"a}fener}, {Vink}, {Kozyreva}, \&
  {Izzard}}]{yoon2012}
{Yoon}, S.-C., {Gr{\"a}fener}, G., {Vink}, J.~S., {Kozyreva}, A., \& {Izzard},
  R.~G. 2012, \aap, 544, L11

\bibitem[{{Yoon} \& {Langer}(2005)}]{yoon2005}
{Yoon}, S.-C. \& {Langer}, N. 2005, \aap, 443, 643

\bibitem[{{Yoon} {et~al.}(2006){Yoon}, {Langer}, \& {Norman}}]{yoon2006}
{Yoon}, S.-C., {Langer}, N., \& {Norman}, C. 2006, \aap, 460, 199

\bibitem[{{Yoon} {et~al.}(2010){Yoon}, {Woosley}, \& {Langer}}]{yoon2010}
{Yoon}, S.-C., {Woosley}, S.~E., \& {Langer}, N. 2010, \apj, 725, 940

\end{thebibliography}

\clearpage

\appendix

\section{Flux correction LH41-1042}\label{sec:correction_appendix}

\begin{figure}[h]
   	\resizebox{\hsize}{!}{\includegraphics{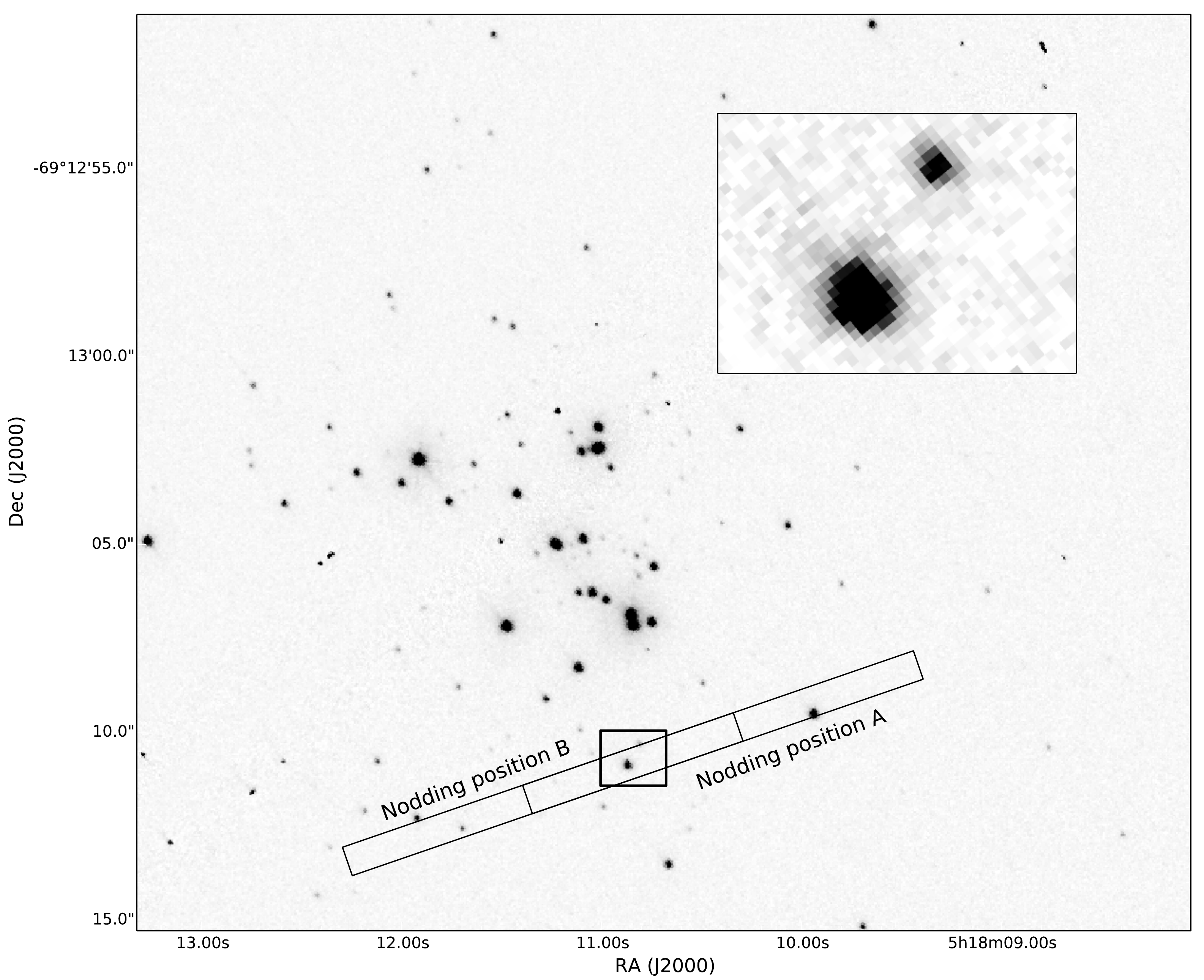}}
  	\caption{Archival HST/WFC3 image of LH41-1042 (F225W filter, proposal ID 12940, PI Massey). North is up and East to the left. The two X-Shooter slit positions are indicated. The inset shows a zoomed image of the boxed area around LH41-1042, and shows the nearby star that may be contaminating the spectrum. }
  	\label{fig:HST}
\end{figure}

As discussed in Section~\ref{sec:obs}, the slope of the spectrum of LH41-1042 before extinction correction is steeper than that of WO star models for an LMC metallicity. This may either be a result of a bad flux calibration, or a second object may be contaminating the spectrum. Inspection of a {\it Hubble Space Telescope} (HST) near-UV image shows that there is indeed a faint star very close to LH41-1042 (Figure~\ref{fig:HST}). We therefore first tried to correct the spectrum by assuming that this object is responsible for the steep slope.

We assume that the flux contribution of the contaminating object follows the Rayleigh-Jeans approximation ($F \propto \lambda^{-4}$) at the X-Shooter wavelength range. This seems justified, as the slope of the uncorrected spectrum is steeper than the slope of the  continuum of an LMC WO-model (which has a free-free emission component and is less-steep than a Rayleigh-Jeans slope) even before dereddening. 

As we cannot determine the reddening from our spectrum, we use the average value for the region found by \cite{massey2005} of $A_V = 0.4$ and a standard total-to-selective extinction of $R_V=3.1$. We assume a luminosity of $1.8\times 10^5 L_{\odot}$ for the WO star, equal to that of BAT99-123. While there is no a-priori reason for the two stars to have the same luminosity, the resulting model flux is in agreement with the dereddened flux in the near-IR region where the contribution of the contaminator becomes negligible.

\begin{figure}
   	\resizebox{\hsize}{!}{\includegraphics{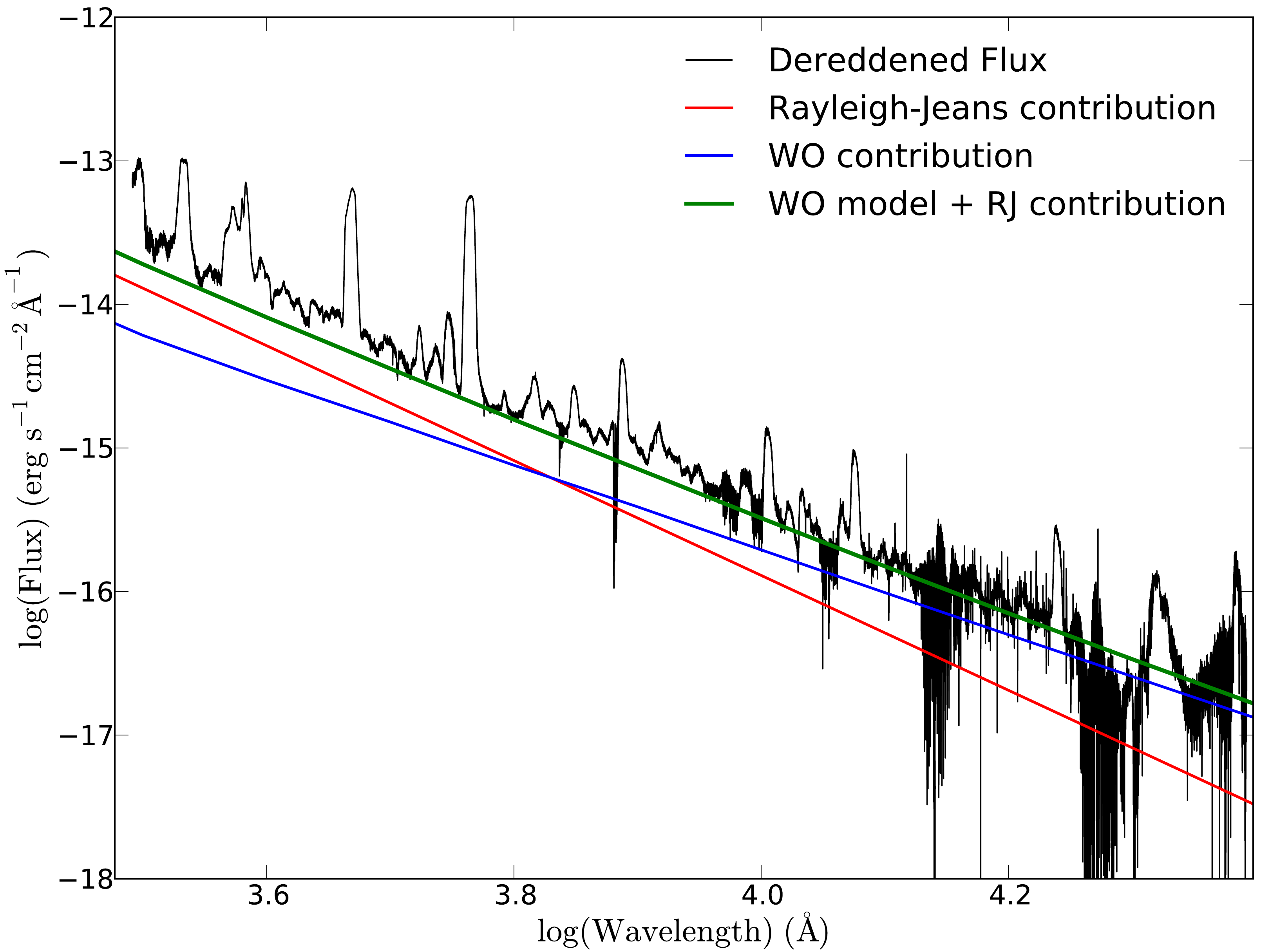}}
  	\caption{Determination of the flux contribution of the companion of LH41-1042. Plotted are the dereddened flux (black), the assumed LMC WO model continuum (blue), the derived Rayleigh-Jeans contribution (red), and the combined continuum from the WO and Rayleigh-Jeans contributions (green).}
  	\label{fig:WO-LMC-2}
\end{figure}

The correction is then done by testing different ratios of the flux contributed by the WO star and the contaminator. Ideally, the flux ratio measured from photometry should be used. However, the only image where the individual stars are resolved is the HST/WFC3 image shown in Figure~\ref{fig:HST}, which uses a filter with an effective wavelength in the near-UV. From this image, the flux ratio is $F_{\mathrm{cont}}/F_{\mathrm{WO}}=0.1$ at 2250\AA. As a much higher flux ratio is needed to recover the observed spectrum, this indicates that the flux of the contaminating object peaks at a wavelength between 2250 \AA \, and the X-Shooter wavelengths (between $\sim 2500-3000$ \AA), and no longer in the Rayleigh-Jeans tail. The contaminating object is therefore likely a B dwarf or giant. This means that the measured flux ratio can not be used to estimate the flux ratio in the X-Shooter wavelength range. Instead, we try out different combinations of flux ratios to obtain a combined spectrum of the WO model and a Rayleigh-Jeans contribution that matches the observed spectrum (Figure~\ref{fig:WO-LMC-2}.)

Using this strategy, we can get a good representation of the slope of the observed spectrum (Figure~\ref{fig:WO-LMC-2}). However, when the Rayleigh-Jeans contribution is subtracted from the observed spectrum, the emission lines become roughly twice as strong as in any of the other WO stars. As otherwise the spectrum of LH41-1042 does not show unusual features, we conclude that this is unlikely to be physical. We therefore assume that the steep slope is a result of the flux calibration, and not of contamination by the faint nearby object. We correct for this by artificially altering the slope to the correct value. While this results in a much more realistic spectrum, the luminosity and mass-loss rate that is derived from the modeling are much more uncertain than those of the other WO stars.

\section{Equivalent widths and spectral classification}\label{sec:SpT_appendix}

\begin{table*}
\centering
\caption{Equivalent width ($W_{\lambda}$) measurements of the lines needed for the spectral classification.}\label{tab:classification}
\begin{tabular}{l c c c c}
\hline\hline \\[-8pt]
ID		& SpT	&$W_{\lambda}$(\ion{O}{vi} $\lambda$3811-34) & $W_{\lambda}$(\ion{O}{v} $\lambda$5590) & $W_{\lambda}$(\ion{C}{iv} $\lambda$5801-12)  \\
	&  & (\AA) & (\AA) & (\AA) \\
\hline \\[-8pt]
WR102		& WO2 		& $1415\pm28$ 	& $196\pm4$		& $156\pm1$\\
WR142 	& WO2		& $1087\pm17$	& $199\pm2$		& $303\pm2$\\
WR93b 	& WO3		& $475\pm5$		& $230\pm4$ 		& $1627\pm7$\\
BAT99-123 	& WO3		& $235\pm4$		& $104\pm1$		& $2115\pm38$\\
LH41-1042 	& WO4		& $171\pm4$		& $172\pm2$ 		& $1869\pm28$\\
DR1 	& WO3		& $146\pm28$		& $125\pm5$		& $417\pm11$\\
\hline
\end{tabular}
\end{table*}

\section{Best-fit models}\label{bestfit_appendix}

\begin{figure*}[h]
   	\resizebox{\hsize}{!}{\includegraphics{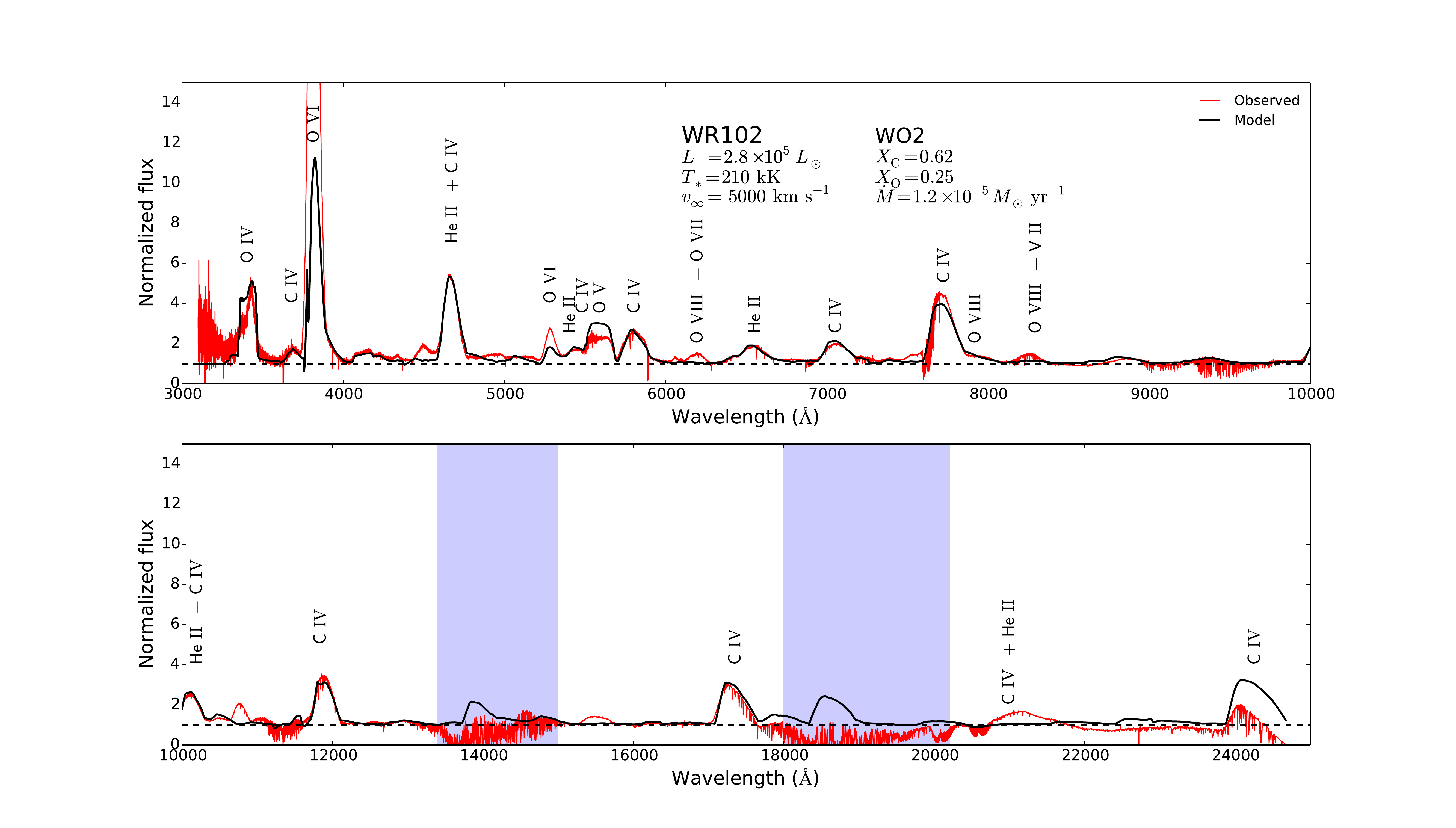}}
  	\caption{Best model of WR102. }
\ \\
   	\resizebox{\hsize}{!}{\includegraphics{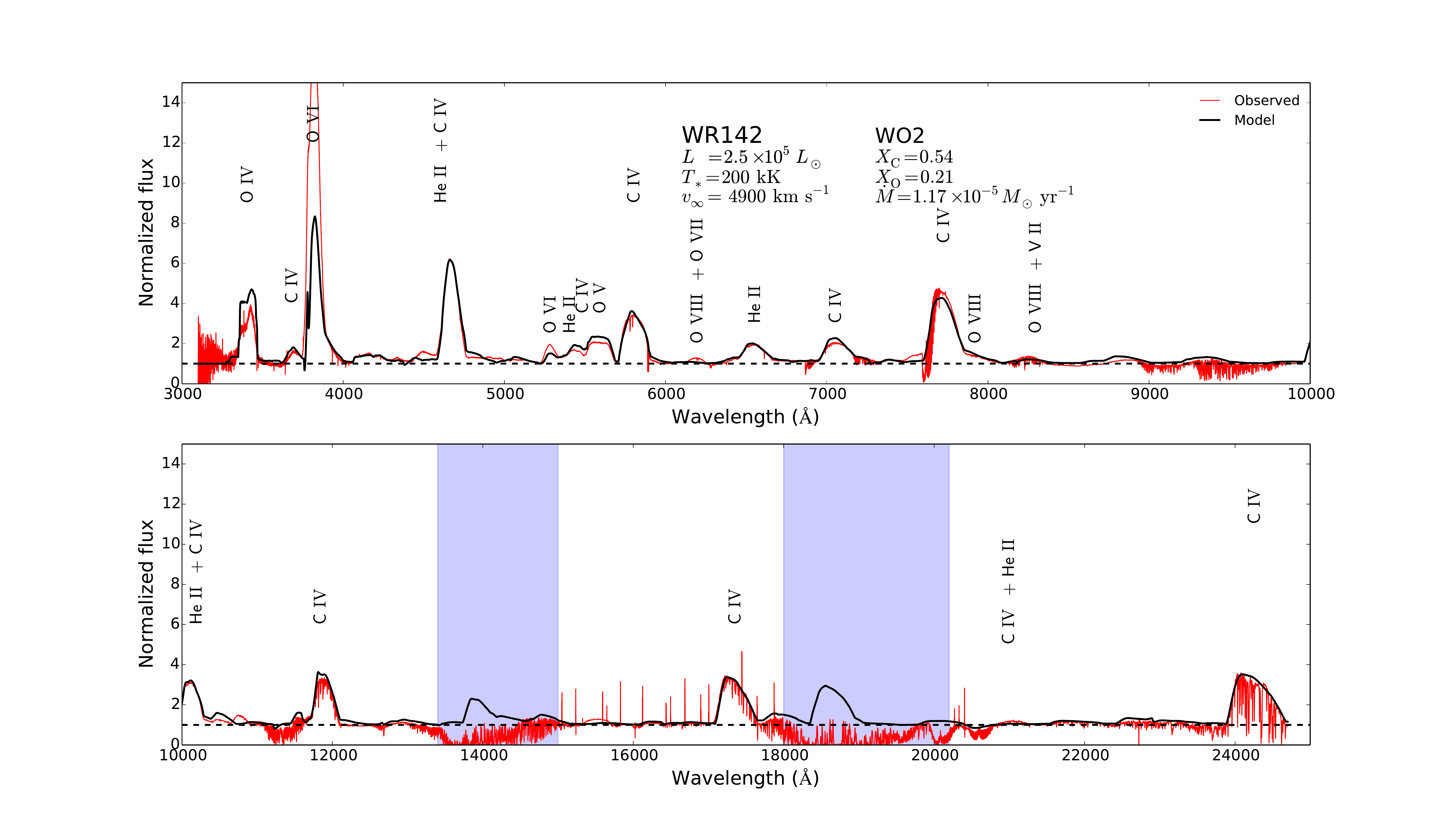}}
  	\caption{Best model of WR142. }
\end{figure*}

\begin{figure*}
   	\resizebox{\hsize}{!}{\includegraphics{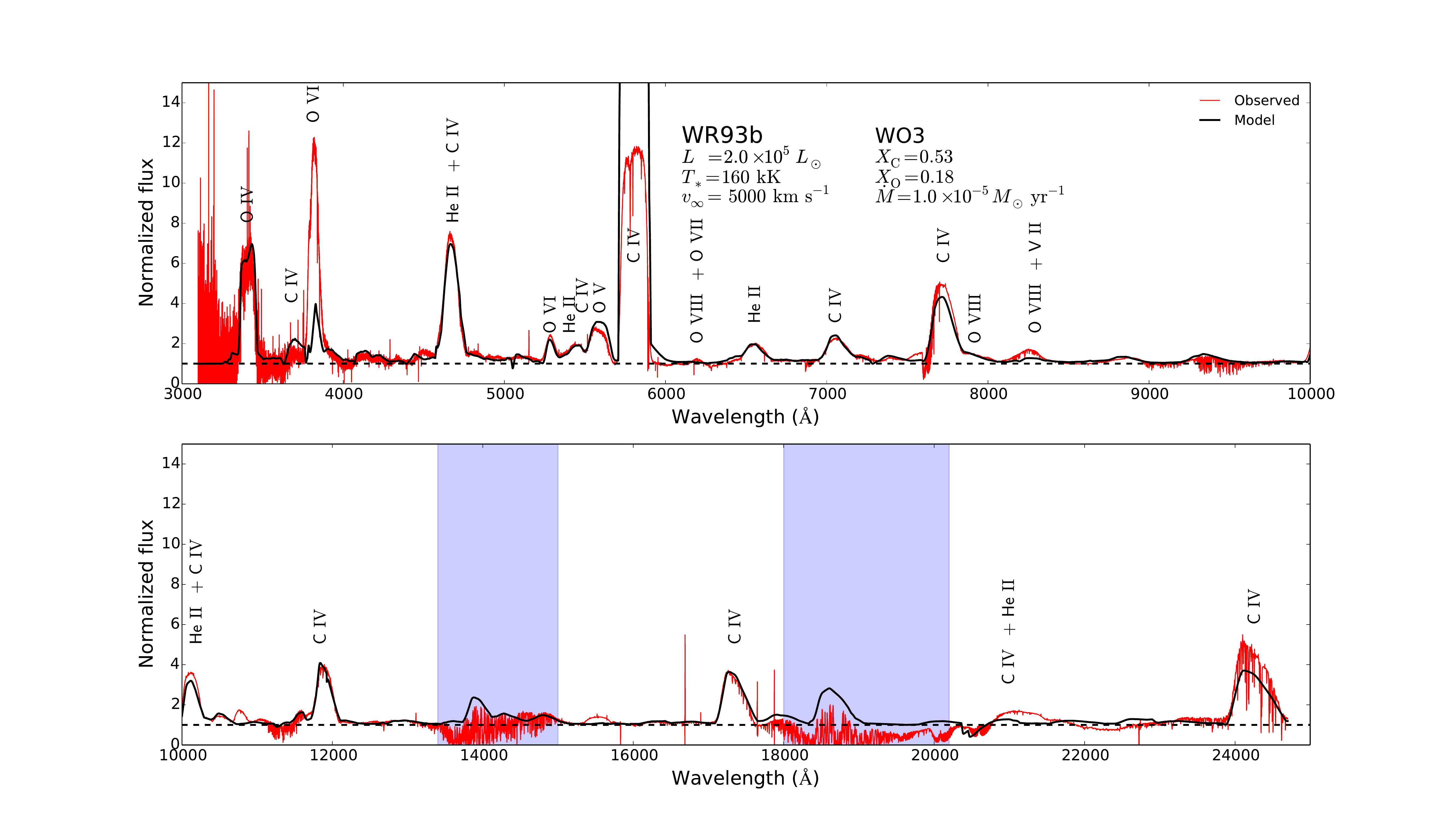}}
  	\caption{Best model of WR93b.}
\ \\
   	\resizebox{\hsize}{!}{\includegraphics{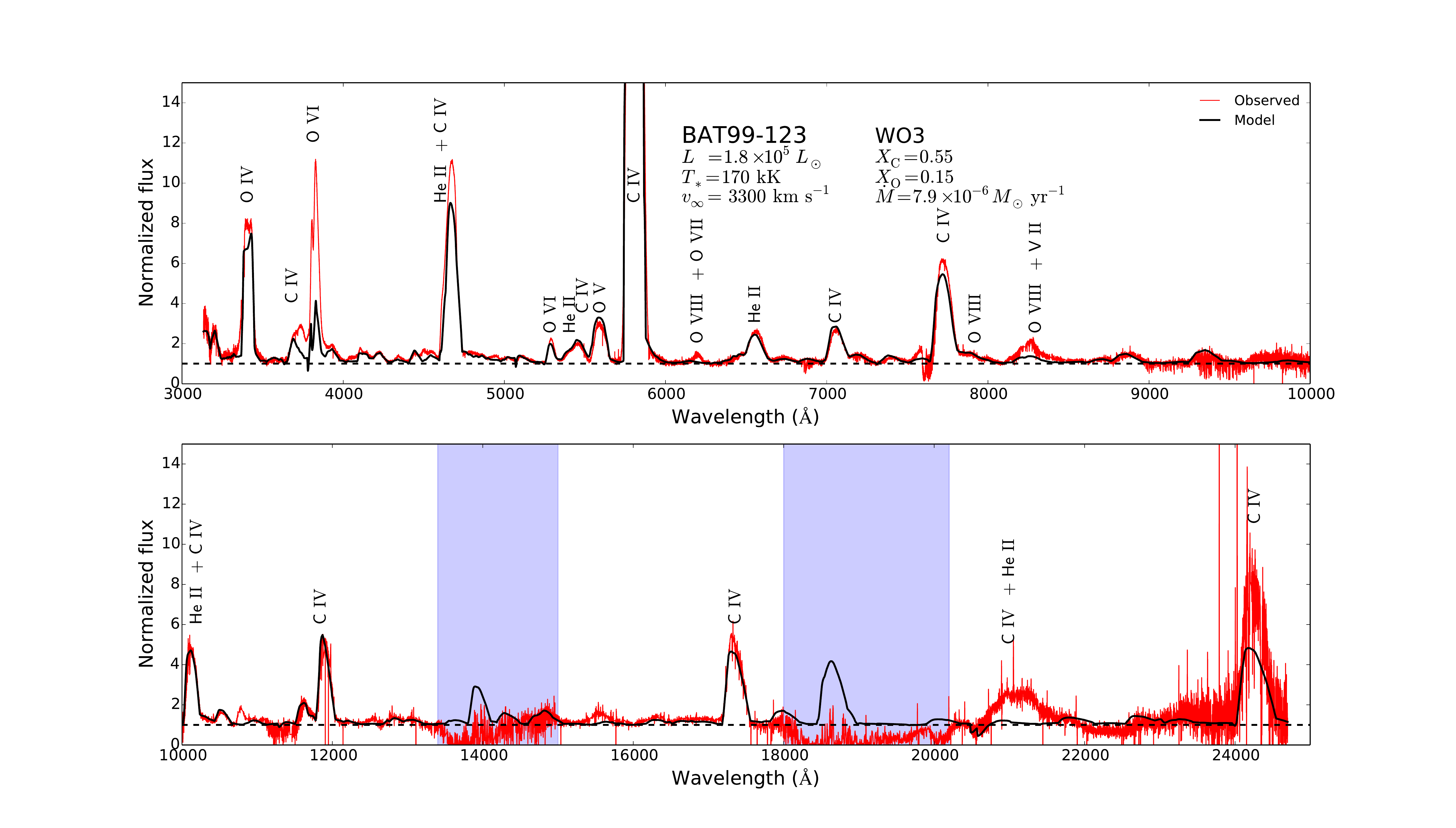}}
  	\caption{Best model of BAT99-123. }
\end{figure*}

\begin{figure*}
   	\resizebox{\hsize}{!}{\includegraphics{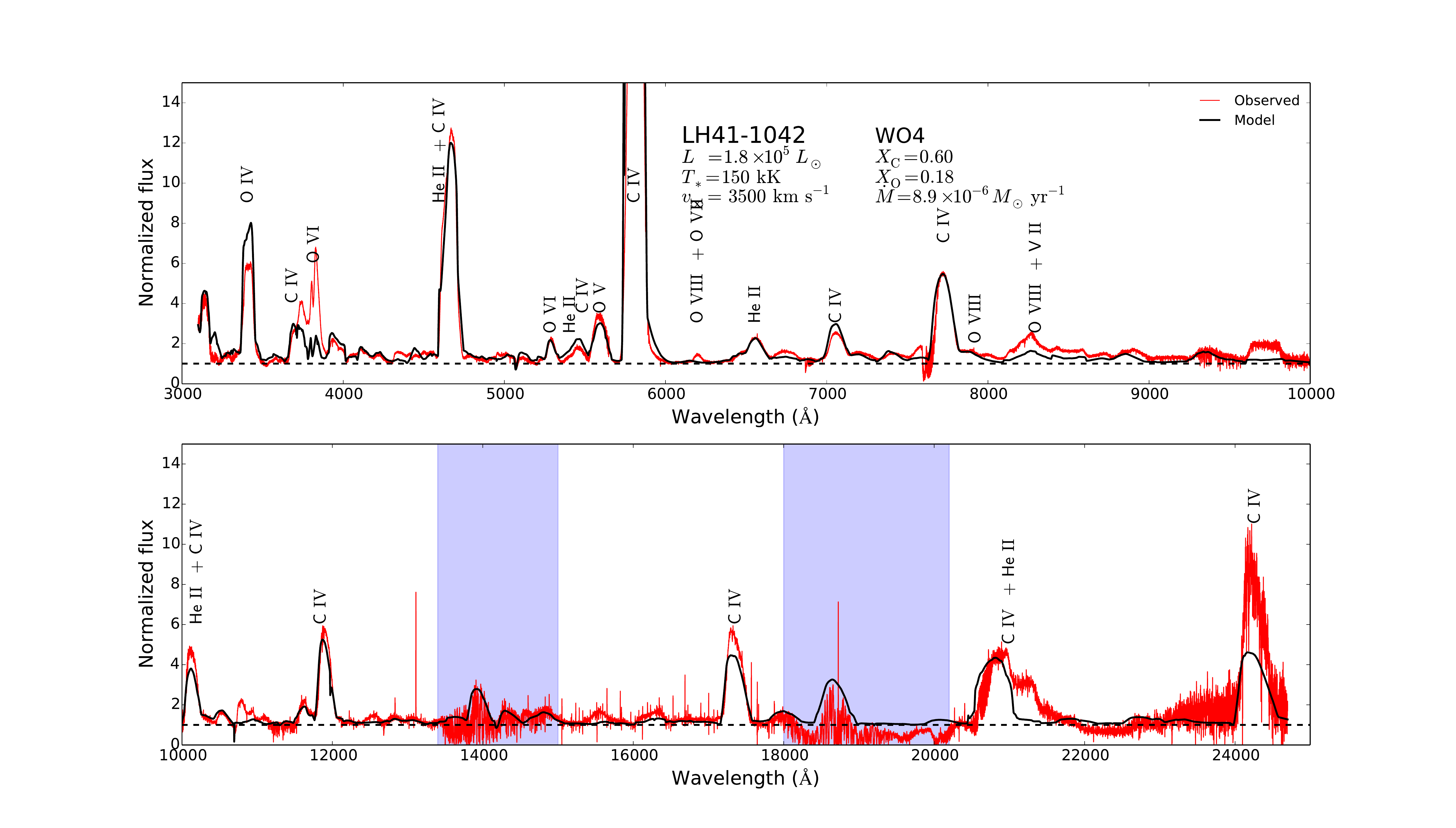}}
  	\caption{Best model of LH41-1042. }
\end{figure*}

\section{Helium-burning models}\label{helium_appendix}

\begin{figure}
   	\resizebox{\hsize}{!}{\includegraphics{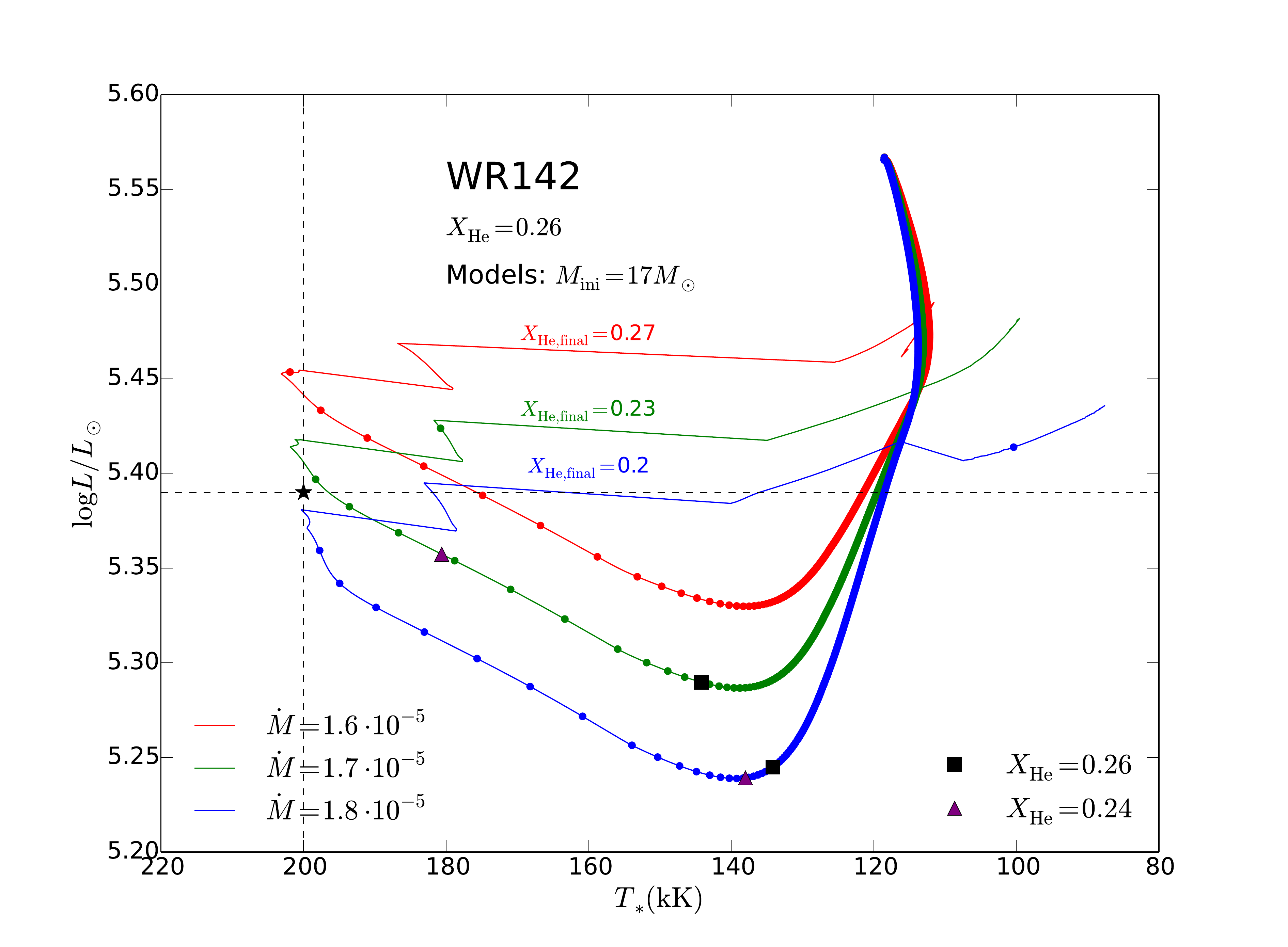}}
  	\caption{Same as Figure~\ref{fig:He_HRD}, but for WR142. }
\end{figure}

\begin{figure}
   	\resizebox{\hsize}{!}{\includegraphics{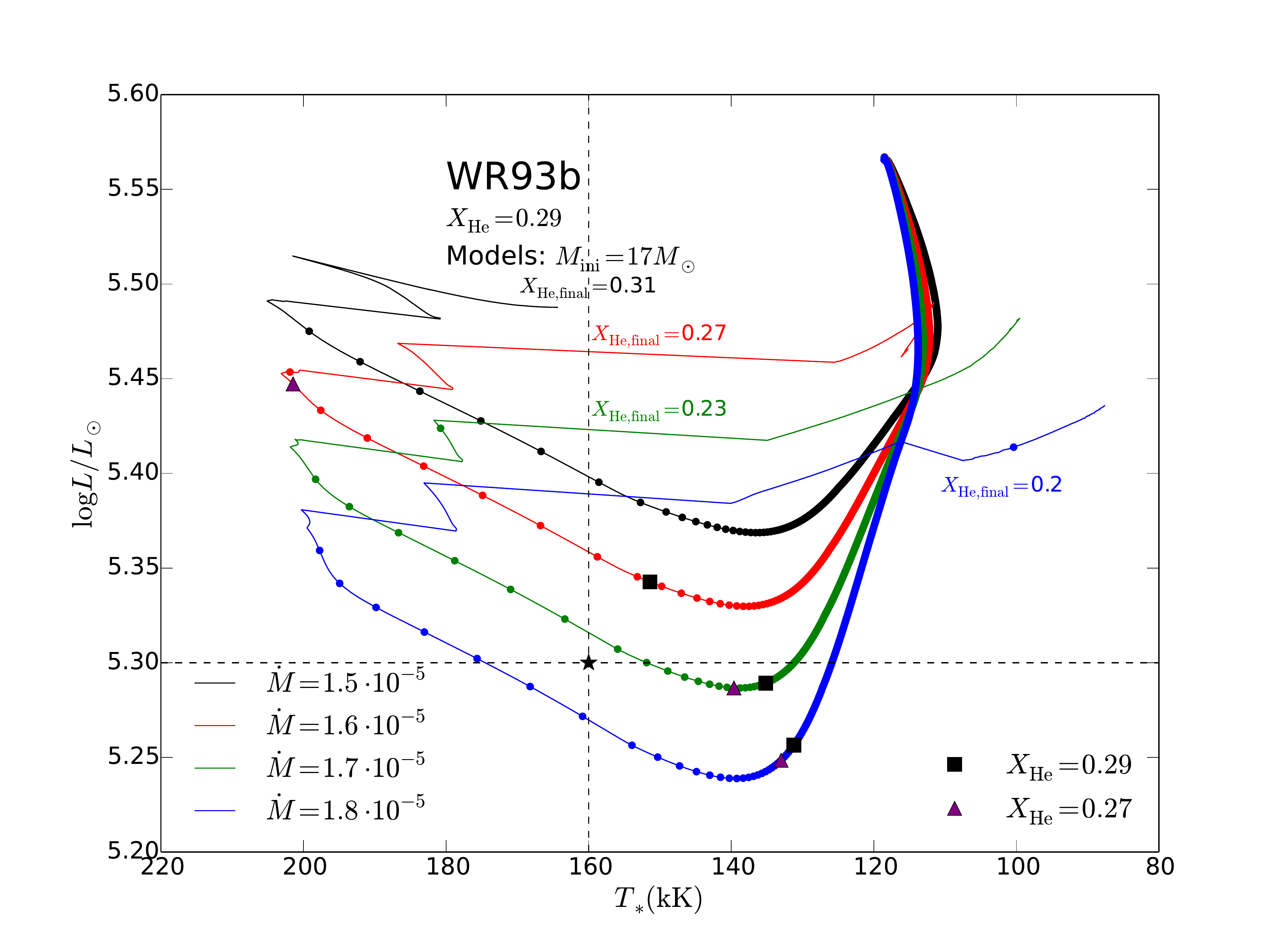}}
  	\caption{Same as Figure~\ref{fig:He_HRD}, but for WR93b. }
\end{figure}

\begin{figure}
   	\resizebox{\hsize}{!}{\includegraphics{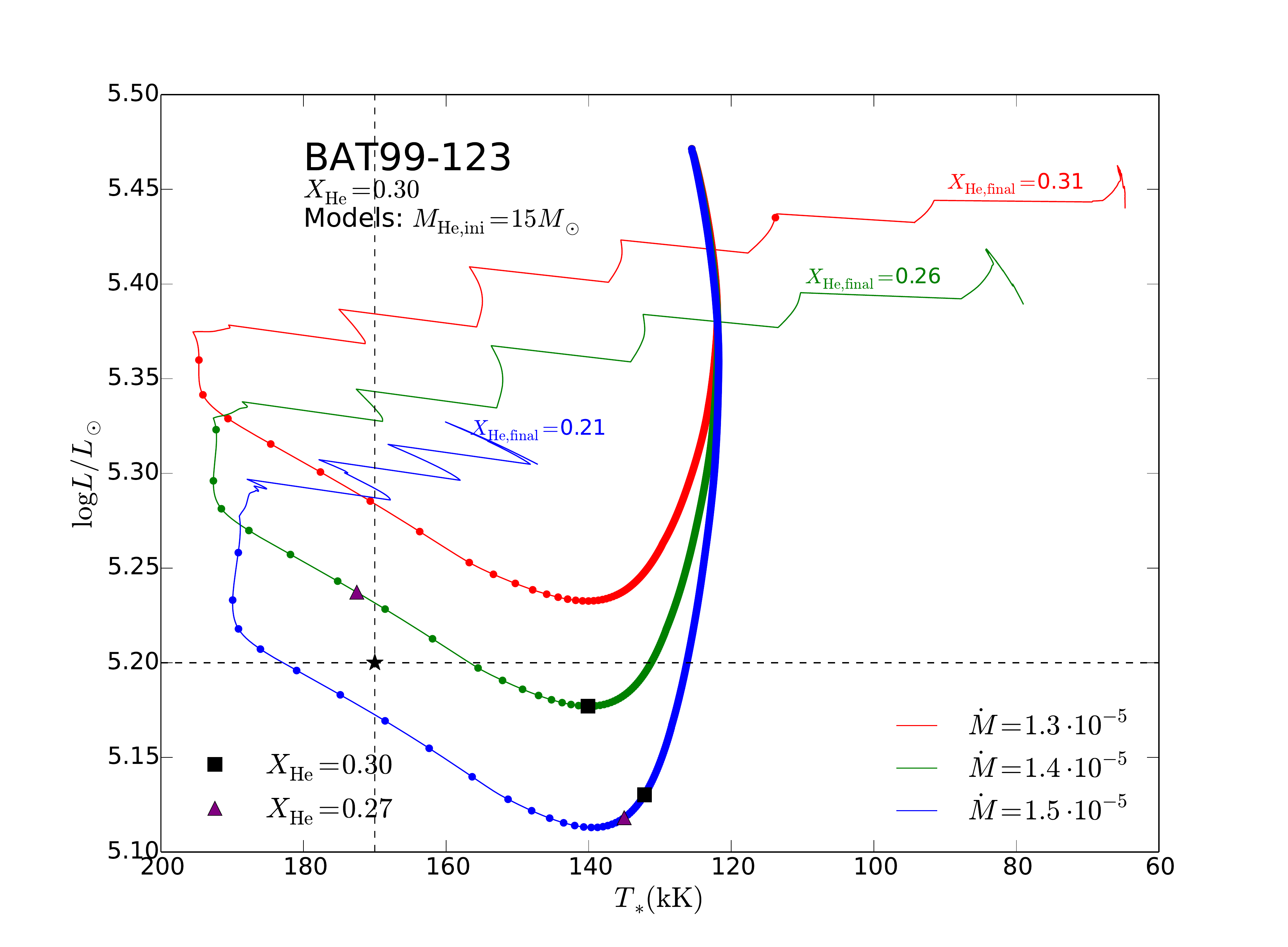}}
  	\caption{Same as Figure~\ref{fig:He_HRD}, but for BAT99-123. }
\end{figure}

\begin{figure}
   	\resizebox{\hsize}{!}{\includegraphics{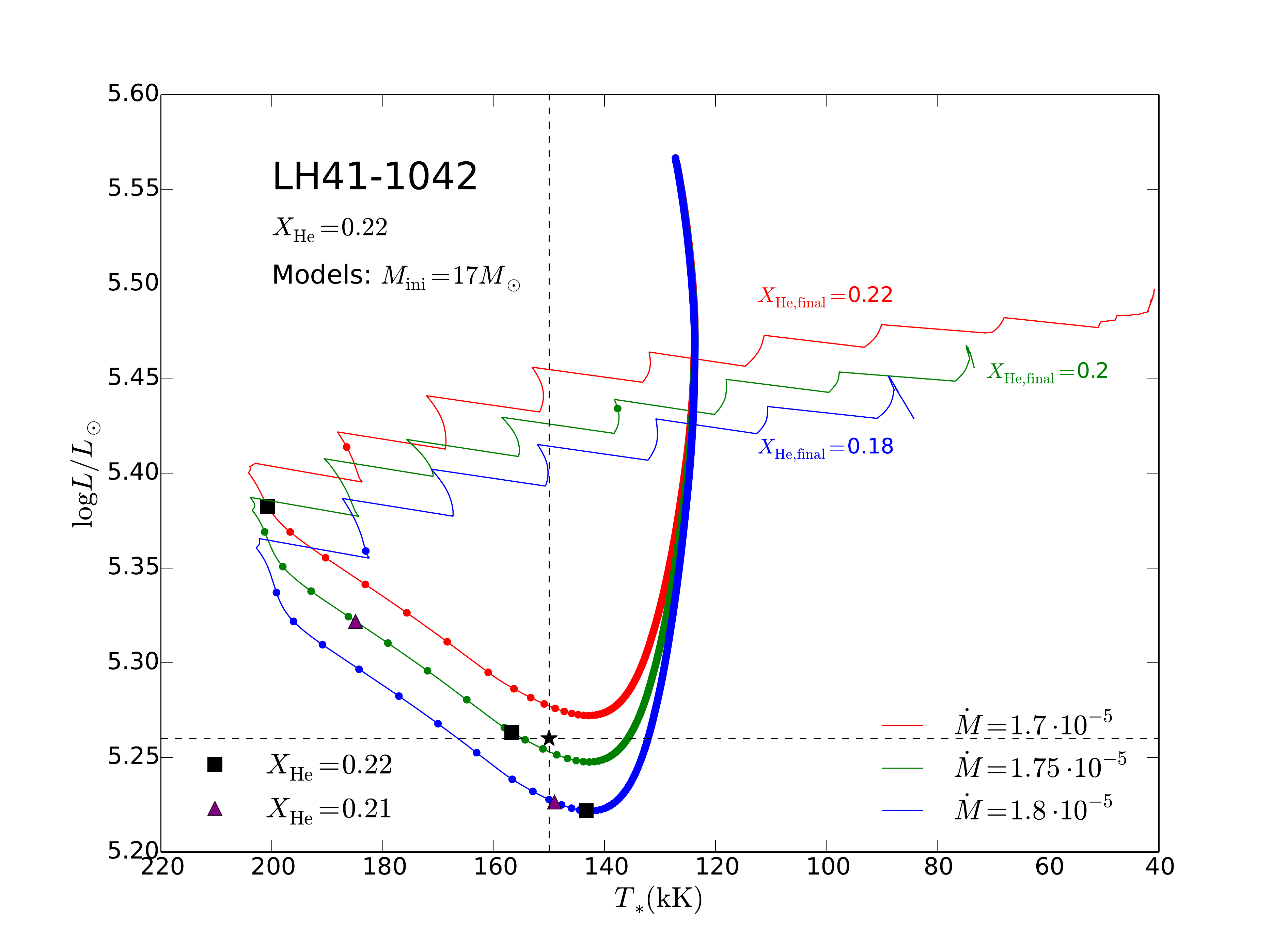}}
  	\caption{Same as Figure~\ref{fig:He_HRD}, but for LH41-1042. }
\end{figure}	

\begin{figure}
   	\resizebox{\hsize}{!}{\includegraphics{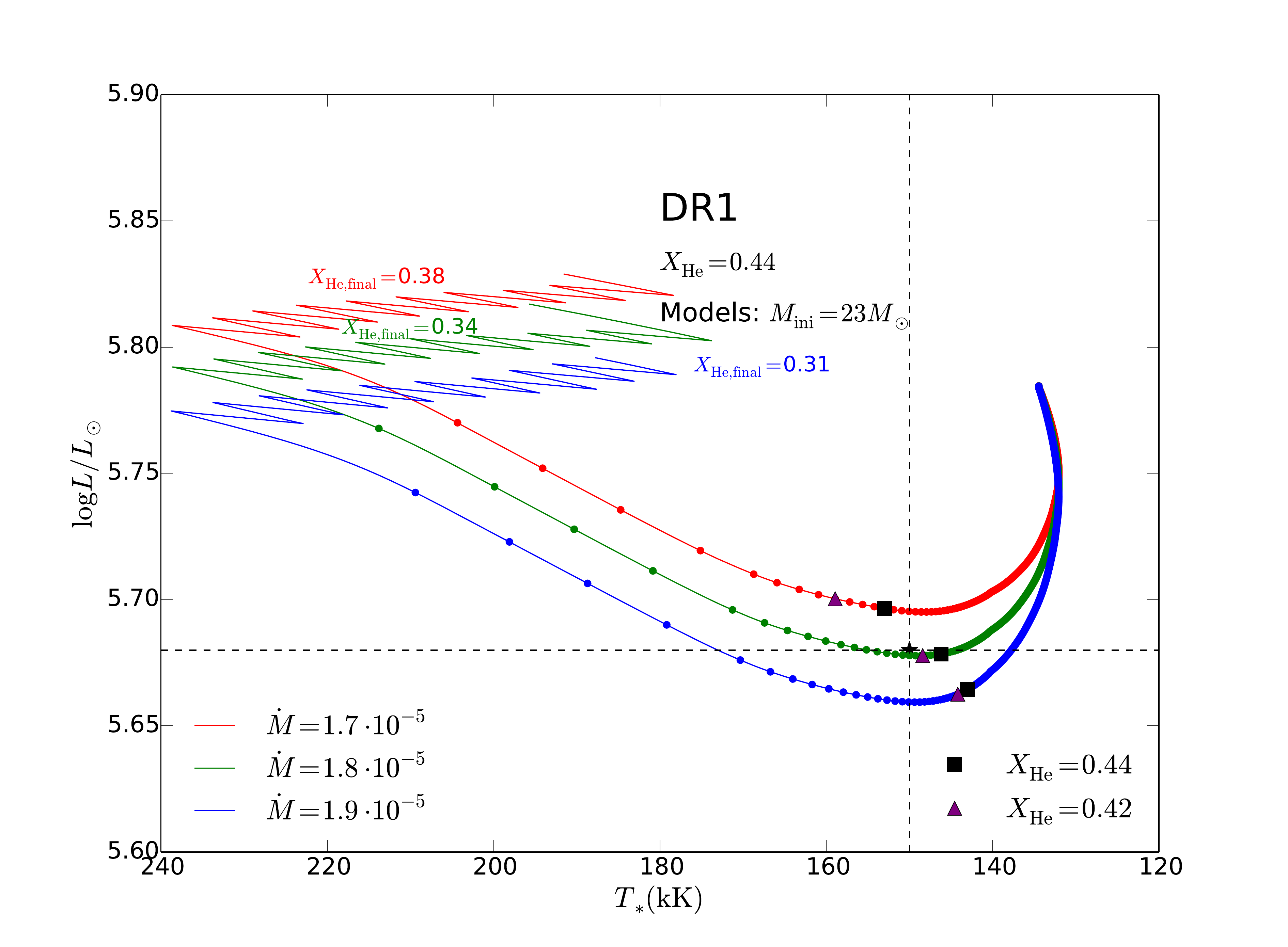}}
  	\caption{Same as Figure~\ref{fig:He_HRD}, but for DR1. }
\end{figure}

\end{document}